\documentclass[paper, new]{geophysics}

\usepackage[scr=rsfs]{mathalpha}

\newcommand{\overds}[1]{\boldsymbol{\mathbf{#1}}} 
\newcommand{\overdv}[1]{\mathbf{#1}} 
\newcommand{\derip}[2]{\frac{\mathrm{\partial}#1}{\mathrm{\partial}#2}} 
\newcommand{\unitary}[1]{\mathbf{\hat{#1}}}

\usepackage{braket}
\usepackage[T1]{fontenc}
\usepackage[utf8]{inputenc}
\usepackage{url}
\usepackage{amsmath, amssymb}
\usepackage[hidelinks]{hyperref}
\usepackage{graphicx}
\usepackage{subfig}
\usepackage{enumitem}
\usepackage{tabularx}
\usepackage{lipsum}
\usepackage{setspace}
\usepackage{xfrac}
\usepackage{siunitx}
\usepackage{multirow}
\usepackage{booktabs}
\usepackage{makecell}
\usepackage{multicol}
\usepackage{xr}
\usepackage{caption}
\usepackage{pdfpages}

\hypersetup{
    pdftitle={Anisotropic rock-scale IP},
    pdfauthor={Charles L. Bérubé},
    bookmarksnumbered=true,
    bookmarksopen=true,
    bookmarksopenlevel=1,
    colorlinks=false,
    allcolors=blue,
    pdfstartview=Fit,
    pdfpagemode=UseOutlines,    
    pdfpagelayout=TwoPageRight
}

\begin{document}

\title{Anisotropic induced polarization modeling with neural networks and effective medium theory}

\renewcommand{\thefootnote}{\fnsymbol{footnote}}

\ms{Submitted to \emph{Geophysics}} 

\address{
\footnotemark[1]Civil, geological and mining engineering department\\
\footnotemark[2]Engineering physics department\\
Polytechnique Montréal\\
C.P. 6079, succ. Centre-ville\\
Montréal QC Canada H3C 3A7\\
\footnotemark[3]Authors contributed equally}
\author{Charles L. Bérubé\footnotemark[1]\footnotemark[3] and Jean-Luc Gagnon\footnotemark[2]\footnotemark[3]}

\footer{Submitted to GEOPHYSICS}
\lefthead{Bérubé \& Gagnon}
\righthead{Anisotropic rock-scale IP}

\maketitle

\includepdf[pages={1}]{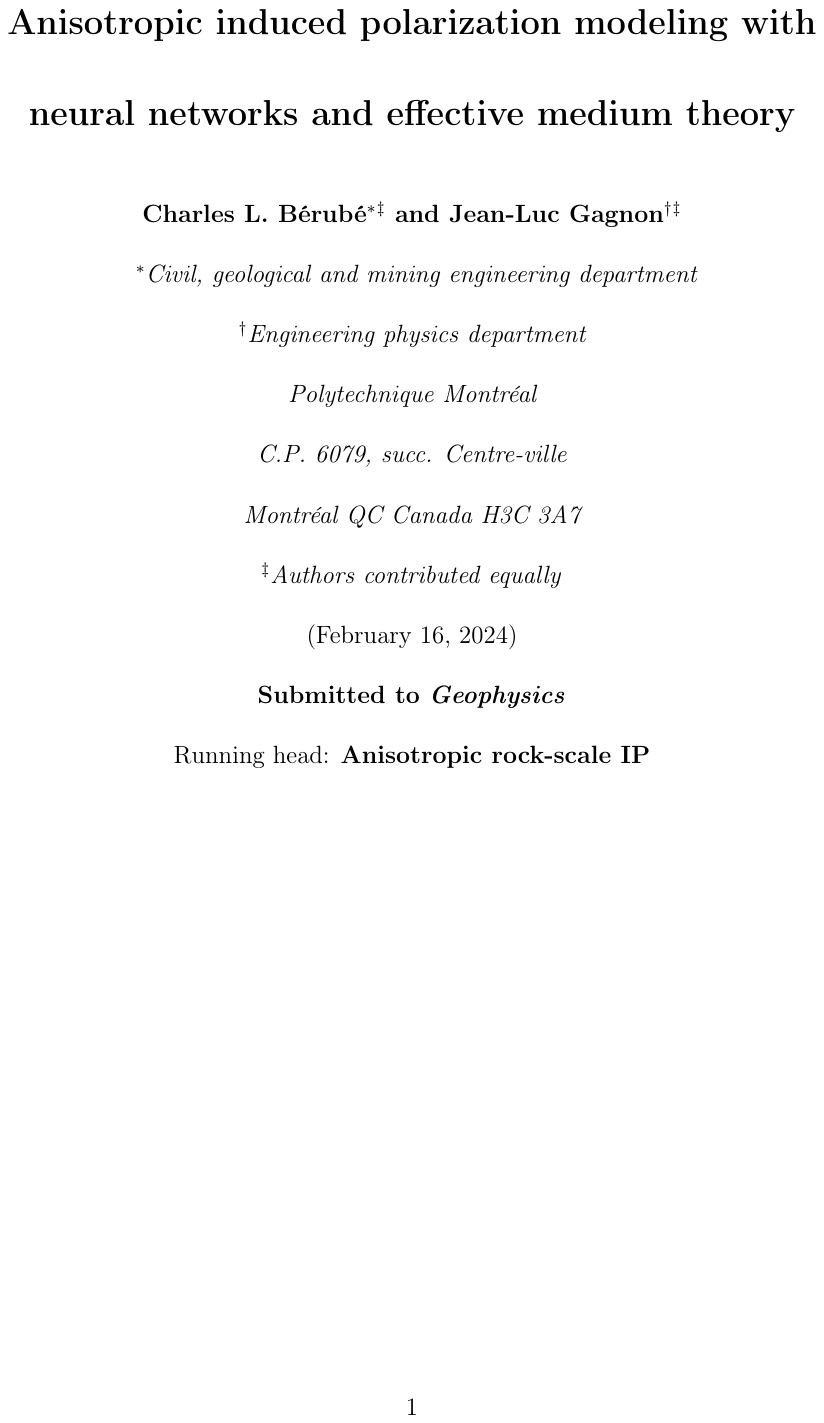}

\begin{abstract}
Accurately interpreting induced polarization (IP) data that reflects the inherent anisotropy of the Earth's crust requires anisotropic IP models. The Generalized Effective Medium Theory of Induced Polarization (GEMTIP) model effectively simulates the IP signatures of rocks containing polarizable minerals. A pivotal element of the GEMTIP model is calculating the depolarization tensor elements, an intensive task for anisotropic rocks because one must numerically solve six parametric integrals for each mineral inclusion. This study aims to streamline anisotropic IP simulations by extending the GEMTIP framework and introducing a machine learning approach to estimate the depolarization tensors. The theoretical contributions of this research are two-fold: (1) we augment the GEMTIP model to encompass anisotropic background conductivity and triaxial ellipsoidal inclusions, and (2) we reformulate the depolarization integrals to normalize their input and output variables, facilitating their estimation by neural networks. The proposed approach eliminates the extensive numerical integration requirements typical of GEMTIP simulations. Validation against analytical solutions for spherical and spheroidal inclusions corroborates the accuracy of the neural network. We then provide examples of IP models for anisotropic rocks containing arbitrarily sized and oriented ellipsoidal inclusions. Analyzing the neural network model, we find that the relationship between chargeability and polarizable inclusion content is increasingly uncertain for increasingly anisotropic rocks. A similar observation applies to the relationship between critical frequency and host rock conductivity. Moreover, the depolarization tensors are, on average, 56~\% sensitive to inclusion anisotropy and 44~\% sensitive to host rock conductivity anisotropy. Remarkably, our neural network drastically accelerates GEMTIP simulations--up to 100,000 times faster than numerical integration--without substantially sacrificing accuracy. This advancement is promising for efficient rock-scale IP modeling in complex and anisotropic geological settings. We release the pre-trained neural network under an open-source Python package as a practical contribution. 
\end{abstract}

\clearpage
\section{Introduction}

The induced polarization (IP) method measures the temporary and reversible energy storage in geomaterials subject to a transient electrical field. Thanks to the myriad of mechanisms contributing to the electrical polarization of rocks and soils, the IP method is increasingly valuable across diverse geoscience and engineering fields. Examples of IP applications include mineral exploration \citep{pelton_mineral_1978, close_electrical_2001, tavakoli_deep_2016, aguilef_relationship_2017, berube_mineralogical_2018, alfouzan_spectral_2020}, mining waste characterization \citep{gunther_spectral_2016, placencia-gomez_pore_2016}, geotechnical engineering \citep{soueid_ahmed_induced_2020}, hydrogeology \citep{gazoty_application_2012, azffri_electrical_2022}, and soil science \citep{schwartz_spectral_2020}.



The primary measure of IP is chargeability, an integrating parameter that describes the dispersive properties of electrical resistivity. The chargeability, $m$, is defined by \cite{seigel_mathematical_1959} as 
\begin{equation}
    m = \frac{\sigma_\infty - \sigma_0}{\sigma_\infty},
\end{equation}
where $\sigma_0$ is the direct current conductivity and $\sigma_\infty$ is the conductivity at infinitely high frequency. However, it is advantageous to formulate the IP phenomenon in terms of frequency-domain effective complex conductivity ($\sigma_\mathrm{eff}$), as in 
\begin{equation}
    \sigma_\mathrm{eff}(\omega) = \sigma'(\omega) + i\sigma''(\omega),
\end{equation}
where $i^2=-1$ is the imaginary unit, $\omega = 2\pi f$, and $f$ is the excitation frequency in Hz. First, the real ($\sigma'$) and imaginary ($\sigma''$) parts of conductivity are straightforward to interpret because they convey the conduction and polarization mechanisms, respectively \citep{binley_resistivity_2020}. Second, the effective conductivity representation simplifies the integration of anisotropy effects \citep{kenkel_2d_2012}. 

Considering the Earth's crust's omnipresent anisotropy, only simulations and analyses that account for anisotropic IP effects can yield accurate data inversions and interpretations in three dimensions. Anisotropy of a rock's effective conductivity is due to (1) fractures, deformations, and foliation resulting in an anisotropic background fabric and (2) elongated polarizable minerals with a preferential orientation within the host rock. Previous research has focused on conductivity anisotropy through experimentation and numerical simulations, yet research on polarization anisotropy remains limited. For example, the anisotropy of argillite samples can result in varied frequency-domain signatures and polarization processes, which poses challenges for developing a comprehensive IP model \citep{cosenza_effects_2007}. Moreover, in permeability estimation applications, the IP method is more sensitive to the anisotropy of sandstone samples than the nuclear magnetic resonance method is \citep{weller_estimating_2010}. While IP data acquisition in orthogonal directions may help resolve the presence of anisotropic host rock \citep{liu_time_2017}, research on anisotropic IP effects caused by elongation or preferential orientation of mineral inclusions is rare. Recent research on the IP signatures of synthetic rocks containing rod-like and sheet-like inclusions shows that chargeability depends on the size and orientation of the inclusions relative to the direction of the polarizing field \citep{gurin_spectral_2021}.

Here, we focus on the frequency-dependent IP effect related to rocks with electron-conducting mineral inclusions. The effective conductivity of such rocks can be modeled phenomenologically through equivalent circuits \citep{pelton_mineral_1978, dias_developments_2000}, mechanistically by solving the Poisson-Nernst-Planck equations \citep{wong_electrochemical_1979, revil_induced_2015, misra_interfacial_2016, bucker_electrochemical_2018, jin_mechanistic_2019} or using a generalization of the classical effective medium theory \citep{zhdanov_generalized_2008, zhdanov_complex_2018}. Adapting effective conductivity models for anisotropic host rocks is typically possible. However, existing models rarely account for the anisotropy of polarizable minerals by assuming that the inclusions are spherical or that the exciting field is perpendicular to rod-like or sheet-like inclusions \citep[e.g.,][]{misra_interfacial_2016}.

\cite{zhdanov_generalized_2008} introduces the Generalized Effective Medium Theory of Induced Polarization (GEMTIP) for simulating the effective conductivity of rocks. It is important to note that the GEMTIP model does not describe IP mechanistically at the scale of pores and grains, but instead at the whole-rock scale according to effective-medium theory \citep{revil_comment_2010}. The GEMTIP model is unique in its ability to simulate the effective conductivity of rocks containing arbitrarily shaped and oriented mineral inclusions and applies to IP data modeling in three dimensions \citep{zhdanov_complex_2018, alfouzan_spectral_2020}. However, GEMTIP simulations require solving the mineral inclusions' volume and surface depolarization tensors. Six depolarization tensor elements must be solved for each ellipsoidal inclusion, and each element is a parametric integral that depends on the lengths of the inclusion's semi-axes and the conductivity of the host medium. On the one hand, the volume depolarization tensor of ellipsoids has no closed-form solution but can be expressed analytically with elliptic integrals of the first and second kinds. On the other hand, solving the surface depolarization tensor of ellipsoidal inclusions requires numerical integration. Both tensors must be solved numerically if the host rock has anisotropic conductivity. Hence, simulating the effective conductivity of actual ore deposits using 3D computer models or petrographic thin section image analysis \citep[e.g.,][]{zhdanov_complex_2018, gurin_induced_2018, berube_mineralogical_2018} and the GEMTIP model is typically subject to prohibitive computation times. Indeed, such rock samples can contain millions of polarizable grains in centimeter-scale samples \citep[e.g.,][]{berube_mineralogical_2018}, which poses a challenge even with the most efficient integration algorithms.

This research investigates using neural networks to streamline GEMTIP simulations considering realistic rock models. Specifically, we aim to accelerate modeling the IP signature of anisotropic rocks by simultaneously solving the six depolarization tensor elements of each mineral inclusion. The Theory section summarizes the development of anisotropic effective conductivity equations. In the Methods section, we normalize the depolarization tensors and describe the machine learning strategy. Finally, the Results section focuses on neural network validation, sensitivity and error analysis, and comparisons with numerical integration methods.

\section{Theory} 

\subsection{Induced polarization effect}
The following assumes that the IP effect is associated with the surface polarization of electron-conducting minerals, mainly caused by the accumulation of electrical charges at the interface between minerals and interstitial water in the host rock. The electrical double layer formed by the accumulation of charges on the surface of grains results in a voltage perturbation $\Delta U$, which is assumed linear and proportional to the current normal to the surface of the grains, such that 
\begin{equation}
    \Delta U = \kappa (\mathbf{\hat{n}}\cdot\mathbf{J}),
    \label{eq:IPeffect}
\end{equation}
where $\mathbf{\hat{n}}$ is a unit vector normal to the grain surface, $\mathbf{J}$ is the current density and
\begin{equation}
    \kappa = \lambda (i\omega)^{-\varrho}
\end{equation}
is a complex-valued and frequency-dependent function with empirical surface polarizability parameter $\lambda$ and relaxation parameter $\varrho$ \citep{zhdanov_anisotropy_2008}. 

\subsection{Green functions and the GEMTIP model}
For any linear differential operator $L$, we can express the solution to the differential equation $L\varphi(\mathbf{r'}) = \psi(\mathbf{r'})$ in terms of an integral representation. Specifically, the solution $\varphi(\mathbf{r'})$, where $\mathbf{r'}$ is an arbitrary reference position, is the volume integral
\begin{equation}
\varphi(\mathbf{r'}) = \int_V G(\mathbf{r},\mathbf{r'})\psi(\mathbf{r'})\,\mathrm{d}\mathbf{r'},
\label{eq:GreenDef}
\end{equation}
where $G(\mathbf{r},\mathbf{r'})$ denotes the Green's function dependent on a position $\mathbf{r}$. The Green's function is uniquely defined by its relationship with $L$, such that
\begin{equation}
\label{eq:Green2}
L (G(\mathbf{r},\mathbf{r'})) = 4\pi\delta(\mathbf{r}-\mathbf{r'}),
\end{equation}
where $\delta$ is the Dirac delta function. 

Given that the operators in Maxwell's equations are linear in free space, using Green's functions is a natural approach for determining the electric field in inhomogeneous matter. \cite{zhdanov_generalized_2008} starts by defining a conductivity tensor $\overds{\sigma}$ as
\begin{equation}
    \overds{\sigma}(\mathbf{r}) = \overds{\sigma}_\mathrm{b} + \boldsymbol{\Delta}\overds{\sigma}(\mathbf{r}),
\end{equation}
where $\boldsymbol{\Delta}\overds{\sigma}(\mathbf{r})$ is a conductivity perturbation and $\overds{\sigma}_\mathrm{b}$ is a constant background conductivity. The quasi-static approximation, ${\partial \mathbf{B}}/{\partial t},{\partial \mathbf{E}}/{\partial t} \approx 0$, implies that
\begin{equation}
    \nabla \cdot (\nabla \times \mathbf{B} + \epsilon_{0}\mu_{0}{\partial\mathbf{E}}/{\partial t}) = \mu_{0}\nabla \cdot \mathbf{J}\implies \nabla \cdot \mathbf{J} = 0, 
\end{equation}
where $\mathbf{B}$ and $\mathbf{E}$ are the magnetic and electric fields, respectively, and where $\epsilon_{0}$ and $\mu_{0}$ are the vacuum permittivity and permeability, respectively. Consequently, the charge conservation law indicates that the volume charge density is constant at the low frequencies typical of IP. Ohm's law, $\mathbf{J} = \overds{\sigma} \mathbf{E}$, then yields 
\begin{align}
\begin{split}
    \nabla \cdot \left(\overds{\sigma}(\mathbf{r})\mathbf{E}(\mathbf{r})\right) &= 0 \\
    \nabla \cdot -\overds{\sigma}_\mathrm{b} \mathbf{E}(\mathbf{r})  &= \nabla \cdot (\boldsymbol{\Delta}\overds{\sigma}(\mathbf{r})\mathbf{E}(\mathbf{r})),
    \label{eq:diffvol}
\end{split}
\end{align}
which is the differential equation governing the electric field inside one inclusion.

The charge accumulation at the boundary between the inclusion and the host, caused by the contrast in medium conductivity and permittivity, implies that the variation of charges on the surface is not zero. Thus, to define the field on the surface of the inclusion, the right side of Equation~\ref{eq:diffvol} must include Maxwell's boundary condition for field components perpendicular to the surface \citep{zhdanov_generalized_2008}. 

When the inclusion size is sufficiently small compared to the scale of the infinite medium, \cite{zhdanov_generalized_2008} defines the relative conductivity tensor $\overds{\xi}$ by
\begin{equation}
    \overds{\xi}(\mathbf{r'}) = \kappa \left(\Delta \overds{\sigma}(\mathbf{r'})\right)^{-1} \overds{\sigma}_\mathrm{b} \overds{\sigma}(\mathbf{r'}) \approx \overds{\xi},
\end{equation}
and the material property tensor $\overds{\chi}$ by 
\begin{align}
    \overds{\chi}(\mathbf{r'}) \cdot \mathbf{E}_{0} = \Delta \overds{\sigma}(\mathbf{r'}) \cdot \mathbf{E}(\mathbf{r'}) \approx \overds{\chi} \cdot \mathbf{E}_{0},
\end{align}
a formulation valid under the quasi-linear approximation because the material property tensor remains constant within the volume of spherical or ellipsoidal inclusions \citep{landau_1984_electrodynamics}. By combining the field on the boundary with the field inside the inclusion to get the total field $\mathbf{E}(\mathbf{r})$, comprised of the surface-induced field $\mathbf{E}_{s}(\mathbf{r})$, the volume-induced field $\mathbf{E}_{v}(\mathbf{r})$ and the volume-averaged field $\mathbf{E}_{0}=\braket{\mathbf{E}(\mathbf{r})}$, and by summing over each inclusion $l$ to get the total field, \cite{zhdanov_generalized_2008} obtains
\begin{align}
\begin{split}
    \mathbf{E}(\mathbf{r}) &= \mathbf{E}_{0} + \mathbf{E}_{s}(\mathbf{r}) + \mathbf{E}_{v}(\mathbf{r})\\
    &= \mathbf{E}_{0} + \sum_{l}\overds{\xi}_{l}\overds{\chi}_{l}\mathbf{E}_{0} \underbrace{\int_{S_{l}}\nabla\nabla' G_{l}(\mathbf{r},\mathbf{r'})\hat{\mathbf{n}_{l}}(\mathbf{r'})\hat{\mathbf{n}_{l}}(\mathbf{r'})\,\mathrm{d}S_l}_{\overds{\Lambda}_{l}} \\ 
    &\qquad + \sum_{l}\overds{\chi}_{l}\mathbf{E}_{0}\underbrace{\int_{V_{l}}\nabla\nabla' G_{l}(\mathbf{r},\mathbf{r'})(\mathbf{r'})\,\mathrm{d}V_{l}}_{\overds{\Gamma}_{l}},
    \label{eq:ChampsTot}
\end{split}
\end{align}
where, $\overds{\Gamma}$ and $\overds{\Lambda}$ are the volume and surface depolarization tensors, respectively. 

The Green's function for the Laplace equation in the low-frequency approximation and when $\overds{\sigma}_\mathrm{b} = \sigma_\mathrm{b}\overdv{I}$, where $\mathbf{I}$ is the identity matrix, is 
\begin{equation}
    G(\mathbf{r},\mathbf{r'}) = \frac{1}{4\pi   \sigma_\mathrm{b}|\mathbf{r}-\mathbf{r'}|}.
    \label{eq:basicLapGreen}
\end{equation}
However, if $\overds{\sigma}_\mathrm{b}$ is an anisotropic diagonal tensor of the form 
\begin{equation}
   \overds{\sigma}_\mathrm{b} = \begin{pmatrix}
    \sigma_{\mathrm{b}, x} & 0 & 0 \\
   0 & \sigma_{\mathrm{b}, y} & 0\\
    0 & 0 & \sigma_{\mathrm{b}, z}
    \label{eq:condNiso}
\end{pmatrix},
\end{equation}
and assuming no applied magnetic field, \cite{stroud_1975_generalized} writes the Green's function as
\begin{equation}
    G(\mathbf{r},\mathbf{r'}) = \frac{1}{4\pi\sigma_{s}}\left(\left(\frac{x-x'}{\sqrt{\sigma_{\mathrm{b}, x}}}\right)^2 + \left(\frac{y-y'}{\sqrt{\sigma_{\mathrm{b}, y}}}\right)^2 + \left(\frac{z-z'}{\sqrt{\sigma_{\mathrm{b}, z}}}\right)^2\right)^{-\frac{1}{2}},
\end{equation}
where $\sigma_{s} = (\sigma_{\mathrm{b}, x}\sigma_{\mathrm{b}, y}\sigma_{\mathrm{b}, z})^{1/2}$, allowing the derivation of a simplified expression that closely resembles Equation~\ref{eq:basicLapGreen} with
\begin{align}
    G(\mathbf{R},\mathbf{R'}) &= \frac{1}{4\pi\sigma_{s}}\left(\left(X - X'\right)^2 + \left(Y - Y'\right)^2 + \left(Z - Z'\right)^2\right)^{-\frac{1}{2}} \nonumber\\
    &= \frac{1}{4\pi\sigma_{s}\left|\mathbf{R}-\mathbf{R'}\right|},
    \label{eq:GreenGen}
\end{align}
using the variable change
\begin{equation}
    \mathbf{R},\mathbf{R'} = \frac{\mathbf{r},\mathbf{r}'}{\boldsymbol{\sigma}_\mathrm{b}^{1/2}},
    \label{eq:TransfoGreen}
\end{equation}
which mirrors the structure of the vectors $\mathbf{r}$ and $\mathbf{r'}$, but where each coordinate is scaled by its corresponding conductivity tensor element \citep{apresyan_depolarization_2014}.

The following sections derive the volume and surface depolarization tensors for triaxial ellipsoidal inclusions in anisotropic host materials. While \cite{zhdanov_anisotropy_2008, zhdanov_complex_2018} address the special case of spheroidal inclusions (namely oblate and prolate ellipsoids), our approach sets itself apart by deriving the integrals for triaxial ellipsoids, by accounting for an anisotropic background conductivity tensor, and by specifically formulating the integrals for approximation via neural networks.

\subsection{Anisotropic volume depolarization tensor}
\label{sect:DepTenVol}
Using a corollary of the divergence theorem, the volume depolarization tensor $\overds{\bf{\Gamma}}_{l}$ of inclusion $l$ in an isotropic host rock simplifies to \citep{stewart_2011_multivariable}
\begin{equation}
    \overds{\bf{\Gamma}}_{l} = \int_{S} \nabla G(\mathbf{r},\mathbf{r'})\hat{\bf{n}}  \, \mathrm{d}S,\label{eq:TensGamma}
\end{equation}
such that 
\begin{align}
    \overds{\Gamma}_{l} &= \int_{\mathcal{S}} \frac{(\bf{r}-\bf{r'})}{4\pi\sigma_\mathrm{b}|\bf{r}-\bf{r'}|^{3}}\frac{\bf{n'}}{|\bf{n'}|}\,\mathrm{d}S \nonumber \\\
     &= -\frac{1}{4\pi\sigma_\mathrm{b}}\int_{S} \frac{\bf{r'}}{|\bf{r'}|^{3}}\frac{\bf{n'}}{|\bf{n'}|}\,\mathrm{d}S.
     \label{eq:partInt}
\end{align}
Note that we evaluate the gradient of the Green's function at $\mathbf{r} = (0,0,0)$, approximating the depolarization tensor of the inclusion as constant throughout its volume. Furthermore, the tensor product $\mathbf{r'}\mathbf{n'}$ stems from the vectors 
\begin{equation}
    \mathbf{r'} = {a}\sin(\theta)\cos(\phi)\,\mathbf{\hat{x}} + {b}\sin(\theta)\sin(\phi)\,\mathbf{\hat{y}} + {c}\cos(\theta)\,\mathbf{\hat{z}}
\end{equation}
and 
\begin{equation}
    \mathbf{n'} = a^{-1}\sin(\theta)\cos(\phi)\,\mathbf{\hat{x}} + b^{-1}\sin(\theta)\sin(\phi)\,\mathbf{\hat{y}} + c^{-1}\cos(\theta)\,\mathbf{\hat{z}},
\end{equation}
where $a$, $b$, and $c$ are the semi-axes lengths of the ellipsoid along the $x$, $y$, and $z$ axes, respectively, and where $\theta$ and $\phi$ are the inclination and azimuth angles, respectively. 

To compute the vector product within the element $\mathrm{d}S = |\mathbf{r_{\theta}}\times \mathbf{r_{\phi}|\,\mathrm{d}\theta\, \mathrm{d}\phi}$, we must parametrize the ellipsoid in its Cartesian form with
\begin{align}
   \mathbf{C}(x,y,z) &= x\,\mathbf{\hat{x}} + y\,\mathbf{\hat{y}} + z\,\mathbf{\hat{z}},
   \label{eq:ellipsoide}
\end{align}
where $x = a\sin(\theta)\cos(\phi)$, $y = b\sin(\theta)\sin(\phi)$, and $z = c\cos(\theta)$.
The ellipsoid parametrization enables the calculation of the general expression for the norm of the vector product, leading to the simplified expression $\mathrm{d}S = abc|\mathbf{n}'|\,\mathrm{d}\theta\,\mathrm{d}\phi$. When assuming $a=b=c$, indicative of a spherical shape, the elements within the depolarization tensor simplify to $(3\sigma_\mathrm{b})^{-1}$. Otherwise, by including the integration bounds, the tensor product and $\mathrm{d}S = |\mathbf{r_{\theta}}\times \mathbf{r_{\phi}|\mathrm{d}\theta \, \mathrm{d}\phi}$ in Equation~\ref{eq:partInt}, the diagonal elements of the volume depolarization tensor are \citep{stewart_2011_multivariable}
\begin{equation}
\overds{\Gamma}_{l} = -\frac{abc}{4\pi\sigma_\mathrm{b}}\int_{0}^{2\pi}\int_{0}^{\pi}\frac{\mathrm{d}\theta \mathrm{d}\phi\sin{(\theta})}{|\mathbf{r}'|^3}\mathbf{n'}\mathbf{r'}.
\label{eq:IsotropicVolumeTensorTheory}
\end{equation}
The non-diagonal elements are periodic integrals or have odd symmetry with respect to $\phi$. By translating the integration bounds, which is permissible due to the periodic nature of the integrands, these elements simplify to zero.

In the presence of anisotropic background conductivity, we must generalize Equation~\ref{eq:IsotropicVolumeTensorTheory} with the modified Green's function in Equation~\ref{eq:GreenGen}. Doing so, we obtain an expression comparable with the isotropic case, but where $\mathbf{r'}$ becomes $\mathbf{R'}$. The variable $\overds{T}$, defined by 
 \begin{equation}
     \overds{T} = \begin{pmatrix}
   \sigma_{\mathrm{b},x} & 0 & 0 \\
   0 & \sigma_{\mathrm{b},y} & 0\\
   0 & 0 & \sigma_{\mathrm{b},z}
    \end{pmatrix}^{-1/2}
 \end{equation}
is also needed in the chain derivation because the gradient is calculated for the Cartesian coordinates. Appendix~A provides the derivation of the anisotropic volume depolarization tensor, which yields 
\begin{equation}
    \overds{\Gamma}_{l} = -\frac{abc}{4\pi\sigma_{s}}\overds{T}\int_{0}^{2\pi}\int_{0}^{\pi}\frac{\mathrm{d}\theta \mathrm{d}\phi\, |\sin{\theta}|}{|\mathbf{R'}|^3}\mathbf{n'}\mathbf{R'},
\label{eq:TensDipEll3}
\end{equation}

\subsection{Anisotropic surface depolarization tensor}
\label{sect:DepTenSurf}
The derivation of the surface depolarization tensor starts by evaluating the gradient of the modified Green's function at $\mathbf{R} = 0$, such that
\begin{align}
\begin{split}
     \overds{{\Lambda}}_{l} &= -\int_{\mathcal{S}} \nabla'\left(\frac{\mathbf{R} - \mathbf{R'}}{4\pi\sigma_s|\mathbf{R} - \mathbf{R'}|^{3}}\right)\frac{\mathbf{n'}}{|\mathbf{n'}|}\frac{\mathbf{n'}}{|\bf{n'}|}\,\mathrm{d}S \\
     &= -\frac{1}{4\pi\sigma_s}\int_{\mathcal{S}} \nabla'\left(\frac{\mathbf{R'}}{|\mathbf{R'}|^{3}}\right)\frac{\mathbf{n'}}{|\mathbf{n'}|}\frac{\mathbf{n'}}{|\bf{n'}|}\,\mathrm{d}S.
     \label{eq:gradSurf}
\end{split}
\end{align} 
The subsequent step involves calculating $\nabla'\left({\bf{R'}}/{|\bf{R'}|^{3}}\right)$ to derive the tensorial expression, for which Appendix~A provides the demonstration. In the case of triaxial ellipsoids and anisotropic conductivity, we obtain a general expression for the surface depolarization tensor integrals, which must be solved numerically, reading
\begin{align}
\begin{split}
    \overds{\Lambda}_{l} &= -\frac{abc}{4\pi\sigma_{s}}\int_{0}^{2\pi}\int_{0}^{\pi}\frac{\mathrm{d}\theta \mathrm{d}\phi\, \sin{\theta}}{|\mathbf{R'}|^5|\mathbf{n}'|}\overds{Q'},
\label{eq:TensDipEll2}
\end{split}
\end{align}
where $\mathbf{Q}' = \left(-3(\mathbf{R'}\mathbf{R'}) + |\mathbf{R'}|^{2}\overds{I}\right)\mathbf{n'}\mathbf{n'}\overds{T}^{2}$.

\subsection{Effective medium conductivity}
The last step in the development of the GEMTIP equations is defining $\overds{\sigma}_{\mathrm{eff}}$, the conductivity of the effective medium, under the quasi-static and quasi-linear approximations. Averaging Ohm's law in the volume, the effective current is
\begin{align}
\begin{split}
     \braket{\mathbf{J}(\mathbf{r'})} &= \braket{\overds{\sigma}_\mathrm{eff}\mathbf{E(\mathbf{r'})}} \\
    &= \overds{\sigma}_\mathrm{b}\braket{\mathbf{E(r'})} + \braket{\mathbf{E(r'})\Delta\overds{\sigma}(\mathbf{r'})} \\
    &= \mathbf{E}_0 (\overds{\sigma}_\mathrm{b} + \braket{\overds{\chi}(\mathbf{r'})}).
    \label{eq:EffCurrent}
\end{split}
\end{align}
We refer the reader to \cite{zhdanov_generalized_2008} for the derivation of the volume-averaged material property tensor $\overds{\chi}$. From Equation~\ref{eq:EffCurrent}, the effective conductivity reads 
\begin{align}
    \overds{\sigma}_{\mathrm{eff}} &=  \overds{\sigma}_\mathrm{b} +\sum_{l=1}^{N}\left[\overds{\mathbf{I}}+\overds{\mathbf{p}}_{l}\right]^{-1}\left[\overds{\mathbf{I}}-\left(\overds{\mathbf{I}}+\overds{\mathbf{p}}_{l}\right) \Delta \overds{\sigma}_{l} \overds{\boldsymbol{\Gamma}}_{l}\right]^{-1}\left[\overds{\mathbf{I}} 
 +\overds{\mathbf{p}}_{l}\right] \Delta \overds{\sigma}_{l} \nu_{l},
 \label{eq:conductLgen}
\end{align}
where $N$ is the total number of inclusions, $\mathbf{p}_l=\overds{\xi}_l\overds{\Gamma}_l^{-1}\overds{\Lambda}_l$ is the surface polarizability tensor and $\nu_l$ is the volumetric fraction of the $l$\textsuperscript{th} inclusion. 
\section{Methods} \label{sect:Methods}

\subsection{Integrand normalization}
We must reformulate the depolarization tensor integrals to have normalized input and output variables to facilitate their approximation by neural networks. First, we normalize $|\mathbf{n}'|$ using anisotropy parameters $A = b/a$ and $B = c/a$ by defining
\begin{align}
\begin{split}
\eta &= \left(\sin^2(\theta)\cos^2(\phi) + A^{-2}\sin^2(\theta)\sin^2(\phi) + B^{-2}\cos^2(\theta)\right)^{1/2}.
  \label{eq:norm1}
\end{split}
\end{align}
Next, to account for anisotropic background conductivity, we introduce two additional anisotropy parameters $C = {\sigma_{\mathrm{b},y}}/{\sigma_{\mathrm{b},x}}$ and $D = {\sigma_{\mathrm{b},z}}/{\sigma_{\mathrm{b},x}}$. We also apply the substitution $A' = b'/a'$ and $B' = c'/a'$, where according to Equation~\ref{eq:TransfoGreen}, $a'=a/{\sigma_{\mathrm{b},x}^{1/2}}$, $b'=b/{\sigma_{\mathrm{b},y}^{1/2}}$, and $c'=c/{\sigma_{\mathrm{b},z}^{1/2}}$, to normalize $|\mathbf{R}'|$ by defining 
\begin{equation}
\begin{split}
\rho_x &= \left(\sin^2(\theta)\cos^2(\phi) + A'^{2}\sin^2(\theta)\sin^2(\phi) + B'^{2}\cos^2(\theta)\right)^{1/2} \\
\rho_y &= \left(A'^{-2}\sin^2(\theta)\cos^2(\phi) + \sin^2(\theta)\sin^2(\phi) + (B'/A')^2\cos^2(\theta)\right)^{1/2}  \\
\rho_z &= \left(B'^{-2}\sin^2(\theta)\cos^2(\phi) + (A'/B')^2\sin^2(\theta)\sin^2(\phi) + \cos^2(\theta)\right)^{1/2},
 \label{eq:norm2}
\end{split}
\end{equation}
and, for convenience, to normalize a modified $|\mathbf{n}'|$ by defining
\begin{align}
\begin{split}
     \eta_{x} &= \sin^2(\theta)\cos^2(\phi) + C^{-1/2}\sin^2(\theta)\sin^2(\phi) + D^{-1/2}\cos^2(\theta) \\
     \eta_{y} &= C^{1/2}\sin^2(\theta)\cos^2(\phi) + \sin^2(\theta)\sin^2(\phi) + (C/D)^{1/2}\cos^2(\theta) \\
     \eta_{z} &= D^{1/2}\sin^2(\theta)\cos^2(\phi) + (D/C)^{1/2}\sin^2(\theta)\sin^2(\phi) + \cos^2(\theta).
      \label{eq:norm3}
\end{split}
\end{align}
Then, reformulating Equations~\ref{eq:TensDipEll3}~and~\ref{eq:TensDipEll2} with Equations~\ref{eq:norm1},~\ref{eq:norm2}~and~\ref{eq:norm3}, the normalized volume depolarization tensor elements are
\begin{align}
\begin{split}
    \Gamma_{x} &= -\frac{abc}{4\pi\sigma_{s}a'^{3}\sigma_{\mathrm{b},x}}\int_{0}^{2\pi}\int_{0}^{\pi}\frac{\mathrm{d}\theta \mathrm{d}\phi\sin{(\theta})}{\rho^3_x}\sin^2(\theta)\cos^2(\phi) \\
   \Gamma_{y} &= -\frac{abc}{4\pi\sigma_{s}a'^{3}\sigma_{\mathrm{b},y}}\int_{0}^{2\pi}\int_{0}^{\pi}\frac{\mathrm{d}\theta \mathrm{d}\phi\sin{(\theta})}{\rho^3_x}\sin^2(\theta)\sin^2(\phi) \\
   \Gamma_{z} &= -\frac{abc}{4\pi\sigma_{s}a'^{3}\sigma_{\mathrm{b},z}}\int_{0}^{2\pi}\int_{0}^{\pi}\frac{\mathrm{d}\theta \mathrm{d}\phi\sin{(\theta})}{\rho^3_x}\cos^2(\theta),
   \label{eq:Intv3}
\end{split}
\end{align}
and the normalized surface depolarization tensor elements are
\begin{align}
\begin{split}
   \Lambda_{x} &= \frac{a^{2}bc}{4\pi\sigma_{s}a'^{5}\sigma_{\mathrm{b},x}^{2}}\int_{0}^{2\pi}\int_{0}^{\pi}\frac{\mathrm{d}\theta \mathrm{d}\phi\, \sin{(\theta)}}{\rho^5_x\eta}\sin^2{(\theta)}\cos^2{(\phi)}\left(-3\eta_{x}+ \rho^2_x \right)\\
    \Lambda_{y} &= \frac{a^{2}bc}{4\pi\sigma_{s}a'^{5}\sigma_{\mathrm{b},y}^{2}}\int_{0}^{2\pi}\int_{0}^{\pi}\frac{\mathrm{d}\theta \mathrm{d}\phi\, \sin{(\theta)}}{\rho^5_x\eta}\sin^2{(\theta)}\sin^2{(\phi)}\left(-3\eta_{y}+ \rho^2_y \right)\\
    \Lambda_{z} &= \frac{a^{2}bc}{4\pi\sigma_{s}a'^{5}\sigma_{\mathrm{b},z}^{2}}\int_{0}^{2\pi}\int_{0}^{\pi}\frac{\mathrm{d}\theta \mathrm{d}\phi\, \sin{(\theta)}}{\rho^5_x\eta}\cos^2{(\theta)}\left(-3\eta_{z}+ \rho^2_z \right).\label{eq:Ints3}
\end{split}
\end{align}
Finally, we define dimensionless depolarization tensors by dividing Equation~\ref{eq:Intv3} by $\mathrm{min}(\sigma_{\mathrm{b},x},\sigma_{\mathrm{b},y},\sigma_{\mathrm{b},z})\times\mathrm{min}(a,b,c)$ and by dividing Equation~\ref{eq:Ints3} by $\mathrm{min}(\sigma_{\mathrm{b},x},\sigma_{\mathrm{b},y},\sigma_{\mathrm{b},z})$. The tensor normalization is easily reversible and further facilitates neural network training by bounding the sum of the tensor elements.

\subsection{Evaluation metrics}
We use two evaluation metrics to compare the actual and predicted depolarization tensor elements. The first metric, $p$, reflects the number correctly predicted significant digits and can be interpreted as element-wise precision. Considering any reference depolarization tensor elements $\Gamma,\Lambda$ and their approximations $\hat{\Gamma},\hat{\Lambda}$, $p$ is
\begin{equation}
    p = -\log_{10}\left|\frac{\hat{\Gamma},\hat{\Lambda}-\Gamma,\Lambda}{\Gamma,\Lambda}\right|.
\end{equation}
The second metric, $r$, can be interpreted as element-wise bias and reads
\begin{equation}
    r = \log_{10}\left(\frac{\hat{\Gamma},\hat{\Lambda}}{\Gamma,\Lambda}\right).
\end{equation}
A distribution of $r$ values that is symmetric, narrow and centered on zero indicates unbiased approximations of the depolarization tensors. 

\subsection{Training and validation data}
We generate $10^6$ sets of anisotropy parameters $(A, B, C, D)\in[0, 1]$ using a four-dimensional Sobol sequence with scrambling \citep[see][]{sobol_distribution_1967, owen_scrambling_1998}. In this study, we use the Sobol sequence generator implemented in the PyTorch library, which employs the direction numbers of \cite{joe_constructing_2008}. Then, for each set in the Sobol sequence, we solve Equations~\ref{eq:Intv3}~and~\ref{eq:Ints3} using Simpson's rule integration (SRI) with $10^6$ evaluation points. As shown in Table~\ref{tab:comp-meth-niter}, $10^6$ points are sufficient to accurately estimate the first seven significant digits of the volume depolarization tensor in the case of spheroidal inclusions and isotropic conductivity. Additional evaluation points do not contribute to precision, but they significantly increase integration time. 

\begin{table}[!ht]
\caption{Approximation quality metrics of the volume depolarization tensor elements using the SRI method with a varying number of evaluation points ($n$). The metrics are obtained by comparing numerical integration results with analytical solutions for spheroidal inclusions. Using $10^6$ evaluation points yields the highest precision ($p$) and lowest bias ($r$) without significantly increasing computation time ($t$).}
\centering
\begin{tabular}{lrrr}
\toprule
$n$ & $p$ & $r$ & $t$ (ms) \\
\midrule
$10^3$ & $3.0 \pm 0.9$ & $(-10 \pm 27) \times 10^{-4}$ & $1.1 \pm 0.1$ \\
$10^4$ & $6.0 \pm 0.7$ & $(5 \pm 8) \times 10^{-6}$ & $3.0 \pm 0.4$ \\
$10^5$ & $7.1 \pm 0.3$ & $(7 \pm 8) \times 10^{-8}$ & $8.7 \pm 0.6$ \\
$10^6$ & $7.2 \pm 0.2$ & $(-3 \pm 2) \times 10^{-8}$ & $31 \pm 3$ \\
$10^7$ & $7.0 \pm 0.2$ & $(5 \pm 2) \times 10^{-8}$ & $280 \pm 10$ \\
\bottomrule
\end{tabular}
\label{tab:comp-meth-niter}
\end{table}

Moreover, Table~\ref{tab:comp-meth-int} justifies our choice of SRI for the integration method. Whereas the methods have similar run times when using $10^6$ evaluation points, the SRI and Boole integration methods outperform Monte Carlo and Trapezoid rule integration in terms of the precision metric $p$. SRI is the only method to correctly predict up to seven significant digits from the analytical volume depolarization tensor solution specific to the case of oblate or prolate ellipsoidal inclusions and isotropic conductivity. We also note that the average bias metric $r$ obtained with SRI is one order of magnitude smaller than that of Boole's method for this test. The computation times in Tables~\ref{tab:comp-meth-niter}~and~\ref{tab:comp-meth-int} are specific to the M1 Pro chip with MPS acceleration enabled by the Torchquad library \citep{gomez_torchquad_2021}.

\begin{table}[!ht]
\caption{Computation time ($t$) and approximation quality metrics $p$ and $r$ for solving the volume depolarization tensors with Simpson, Boole, Trapezoid and Monte Carlo methods using $10^6$ evaluation points. The metrics are obtained by comparing numerical integration results with analytical solutions for spheroidal inclusions.}
\centering
\begin{tabular}{lrrr}
\toprule
Method & $p$ & $r$ & $t$ (ms) \\
\midrule
Boole & $6.7 \pm 0.1$ & $(9 \pm 2) \times 10^{-8}$ & $25 \pm 1$ \\
Monte Carlo & $3.2 \pm 0.3$ & $(7 \pm 16) \times 10^{-5}$ & $17 \pm 1$ \\
Simpson & $7.1 \pm 0.1$ & $(8 \pm 14) \times 10^{-9}$ & $25 \pm 5$ \\
Trapezoid & $5.53 \pm 0.07$ & $(-2 \pm 1) \times 10^{-6}$ & $23 \pm 1$ \\
\bottomrule
\end{tabular}
\label{tab:comp-meth-int}
\end{table}

Finally, we concatenate the depolarization tensors calculated with SRI as reference values $\mathbf{\Lambda}^\frown\mathbf{\Gamma}$ for the neural network training. We then split the Sobol sequence into a training dataset denoted by $\mathcal{D}_\mathrm{t}$ and a validation dataset denoted by $\mathcal{D}_\mathrm{v}$. Specifically, $\mathcal{D}_\mathrm{t}$ contains $8\times 10^5$ sets of anisotropy parameters and their corresponding depolarization tensors, whereas $\mathcal{D}_\mathrm{v}$ contains the remaining $2\times 10^5$ sets.

\subsection{Neural network approximation}
The proposed neural network is a multilayer perceptron (MLP) with $K$ hidden layers. The MLP aims to approximate a concatenation of the volume and surface depolarization tensors, denoted by $\mathbf{\Lambda}^\frown\mathbf{\Gamma}$, given a set of anisotropy parameters $(A, B, C, D)$. The output of the MLP is the approximation $\mathbf{\hat{\Lambda}}^\frown\mathbf{\hat{\Gamma}}$, which we define as
\begin{equation}
    \mathbf{\hat{\Lambda}}^\frown\mathbf{\hat{\Gamma}} = \left(\mathbf{W}^{(K+1)}\mathbf{a}^{(K)} + \mathbf{b}^{(K+1)}\right),
\end{equation}
where the output of the $k$\textsuperscript{th} hidden layer is
\begin{equation}
    \mathbf{a}^{(k)} = \operatorname{SiLU}\left(\mathbf{W}^{(k)}\mathbf{a}^{(k-1)} + \mathbf{b}^{(k)}\right), 
\end{equation}
and where 
\begin{equation}
    \operatorname{SiLU}(\cdot) = \frac{(\cdot)}{1+e^{-(\cdot)}}\,
\end{equation}
is the sigmoid linear unit activation function. $\mathbf{W}^{(k)}$ and $\mathbf{b}^{(k)}$ are respectively the weight matrices and bias vectors to optimize. The input layer is a special case with $\mathbf{a}^{(0)} = (A, B, C, D)$. In this work, $K=4$ and hidden layers have a dimension of 128.

\begin{figure}[!htp]
\centering
\includegraphics[width=0.5\textwidth]{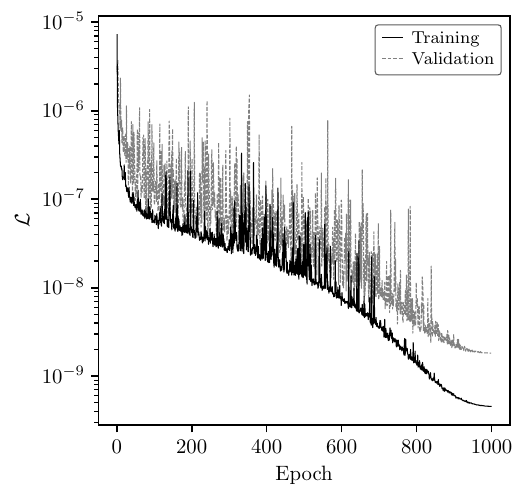}
\caption{MLP training and validation losses ($\mathcal{L}$) as a function of training epoch.}
\label{fig:learning-curves}
\end{figure}

We optimize $\mathbf{W}^{(k)}$ and $\mathbf{b}^{(k)}$ on 1000 epochs of $\mathcal{D}_\mathrm{t}$ using the backward propagation algorithm and the Adam optimizer \citep{kingma_adam_2015}, a batch size of 32 and a maximum learning rate of $10^{-2}$. During training, the learning rate is adjusted according to the one-cycle strategy of \cite{smith_super-convergence_2019}. The loss function ($\mathcal{L}$) we use to optimize the neural network reads
\begin{equation}
\mathcal{L}=\lVert \mathbf{\hat{\Lambda}}^\frown\mathbf{\hat{\Gamma}} - \mathbf{\Lambda}^\frown\mathbf{\Gamma} \rVert_2^2,
\label{eq:lossfunction}
\end{equation}
which is the mean square error between the MLP and SRI approximations of the depolarization tensors. After each optimization step, we obtain a validation loss by computing $\mathcal{L}$ across $\mathcal{D}_\mathrm{v}$ to prevent overfitting. Figure~\ref{fig:learning-curves} shows the neural network training and validation losses as a function of the number of training epochs. 

\subsection{Rotation of the inclusions}
To get a realistic effective rock medium, it is possible to simulate mineral inclusions with random or preferential orientations by rotating the depolarization tensors according to Euler angles $\alpha$, $\beta,$ and $\gamma$. We use the right-hand rule extrinsic rotation matrix convention where $\alpha$ is the angle about the $x$ axis, $\beta$ is about the $y$ axis, and $\gamma$ is about the $z$ axis. The rotated depolarization tensors $\overdv{\Gamma}_\mathrm{rot}$ and $\overdv{\Lambda}_\mathrm{rot}$ are
\begin{equation}
\overdv{\Gamma}_\mathrm{rot}, \overdv{\Lambda}_\mathrm{rot} = \mathbf{S} \overdv{\Gamma} \mathbf{S}^\intercal, \mathbf{S} \overdv{\Lambda} \mathbf{S}^\intercal,
\end{equation}
where 
\begin{equation}
\mathbf{S} = \begin{bmatrix}
        \cos\beta\cos\gamma &
          \sin\alpha\sin\beta\cos\gamma - \sin\alpha\cos\gamma &
          \cos\alpha\sin\beta\cos\gamma + \sin\alpha\sin\gamma \\
        \cos\beta\sin\gamma &
          \sin\alpha\sin\beta\sin\gamma + \cos\alpha\cos\gamma &
          \cos\alpha\sin\beta\sin\gamma - \sin\alpha\cos\gamma \\
       -\sin\beta & \sin\alpha\cos\beta & \cos\alpha\cos\beta \\
         \end{bmatrix}.
\end{equation}

\section{Results}


\subsection{Neural network validation}
This section validates the accuracy of the trained MLP by evaluating its approximation quality metrics and by comparing its effective conductivity predictions against that of simplified media for which analytical solutions exist.

\subsubsection{Depolarization tensor elements}
We start by validating the neural network with a blind test procedure. The test consists of using all sets of anisotropy parameters $(A, B, C, D)$ from $\mathcal{D}_\mathrm{v}$ as inputs for the MLP to predict their corresponding volume and surface depolarization tensor elements. We then compare the MLP outputs with their SRI counterparts to evaluate the $p$ and $r$ metrics, averaging them over $\mathcal{D}_\mathrm{v}$. Table~\ref{tab:validation-metrics} summarizes each depolarization tensor element's $p$ and $r$ validation metrics. 

\begin{table}[!ht]
\caption{Validation metrics for each depolarization tensor element. The values and uncertainties are, respectively, the mean and standard deviation of the mean for $\mathcal{D}_\mathrm{v}$.}
\centering
\begin{tabular}{lrrr}
\toprule
{} & {$p$} & {$r$ $(10^{-3})$} \\
\midrule
$\Gamma_{x}$ & $3.118 \pm 0.001$ & $-0.56 \pm 0.05$ \\
$\Gamma_{y}$ & $3.556 \pm 0.001$ & $0.03 \pm 0.02$ \\
$\Gamma_{z}$ & $3.582 \pm 0.001$ & $-0.03 \pm 0.02$ \\
$\Lambda_{x}$ & $2.869 \pm 0.001$ & $-1.9 \pm 0.1$ \\
$\Lambda_{y}$ & $3.385 \pm 0.001$ & $0.74 \pm 0.04$ \\
$\Lambda_{z}$ & $3.362 \pm 0.001$ & $0.39 \pm 0.05$ \\
Average & $3.312 \pm 0.001$ & $-0.23 \pm 0.05$ \\
\bottomrule
\end{tabular}
\label{tab:validation-metrics}
\end{table}

The blind test shows the MLP predicts the depolarization tensor elements to three significant digits on average ($p$ metric in Table~\ref{tab:validation-metrics}). Approximations of $\Gamma_{x}$ and $\Lambda_{x}$ are marginally less accurate due to numerical instability of the integrands for $A$ and $B$ values below 0.1. The $r$ metric indicates low spread and bias in predictions, except for $\Lambda_{x}$, which has a notably higher $r$ value. While increasing MLP complexity, training time and training data could enhance accuracy, three significant digits are deemed sufficient. Indeed, further precision in predicting the integrals is unnecessary for reproducing known analytical solutions, as demonstrated in the next section.

\subsubsection{Complex conductivity of spheroidal inclusions}

Here, we validate the MLP by predicting the effective conductivity of rocks containing relatively simple inclusion geometries, such as spheres and rotational ellipsoids, and comparing the predictions with those of previously published work. 

\begin{figure}[!htp]
\centering
\includegraphics[width=0.5\textwidth]{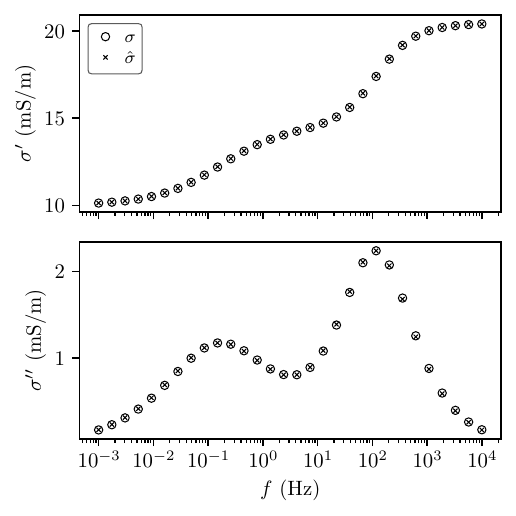}
\caption{Validation of the neural network effective conductivity ($\hat{\sigma}$) for a rock with spherical inclusions against the analytical solution ($\sigma$) of \cite{zhdanov_generalized_2008}.}
\label{fig:validation-sphere}
\end{figure}

Regarding spherical inclusions, analytical solutions to the depolarization tensor elements exist and are extensively discussed in \cite{zhdanov_generalized_2008}. Here, we consider a mixture of spherical inclusions in an isotropic background, i.e., $(A,B,C,D)=1$. Moreover, we divide the inclusions into two groups with contrasting volumetric fractions, sizes, and physical properties by setting $\sigma_\mathrm{b}=0.01$~S/m, $a_l=[0.1, 0.2$]~mm, $\nu_l=[0.20, 0.15]$, $\sigma_l=[10, 1000]$~S/m, $\varrho_l=[0.8, 0.6]$, and $\lambda_l=[1, 0.01]$~$\Omega$m$^2/$s$^{\varrho_l}$. Figure~\ref{fig:validation-sphere} compares the MLP-predicted $\overds{\sigma}_\mathrm{eff}$ of this mixture against its analytical solution.

\begin{figure}[!htp]
\centering
\includegraphics[width=0.5\textwidth]{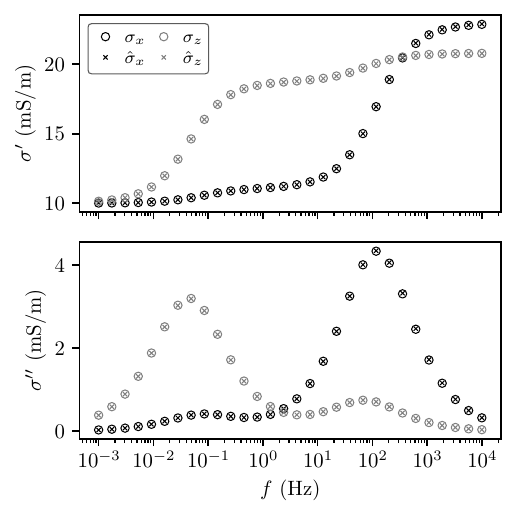}
\caption{Validation of the neural network effective conductivity ($\hat{\sigma}$) for a rock with spheroidal inclusions against the proposed solution ($\sigma$) of \cite{zhdanov_complex_2018}.}
\label{fig:validation-ellipsoid}
\end{figure}

Regarding spheroidal inclusions, analytical solutions for the volume depolarization tensor and numerical integrals for the surface depolarization tensor are provided in \cite{zhdanov_complex_2018}. Here, we use two types of spheroidal inclusions: one oblate ($A=1$, $B=0.2$) and one prolate ($A,B=0.2$). The background conductivity is $\sigma_\mathrm{b}=0.01$~S/m and is isotropic ($C,D=1$). The oblate and prolate inclusions have contrasting volumetric fractions and sizes, but identical physical properties: $a_l=[0.01, 10$]~mm, $\nu_l=[0.15, 0.05]$, $\sigma_l=[0.1, 0.1]$~S/m, $\varrho_l=[0.8, 0.8]$ and $\lambda_l=[0.5, 0.5]$~$\Omega$m$^2/$s$^{\varrho_l}$. Figure~\ref{fig:validation-ellipsoid} compares the MLP-predicted $\overds{\sigma}_\mathrm{eff}$ of this mixture to the solutions of \cite{zhdanov_complex_2018}. Here, the minor axis of the oblate spheroids and the major axis of the prolate spheroids align with the $z$-axis. Consequently, the polarization anisotropy manifests itself in the form of increased $\sigma''$ in the $z$ direction for the prolate spheroids (grey markers in Figure~\ref{fig:validation-ellipsoid}) and increased $\sigma''$ in the $x$ direction for oblate spheroids (black markers in Figure~\ref{fig:validation-ellipsoid}). Conductivity in the $y$ direction is equal to that in the $x$ direction for both inclusion types and is not illustrated.

Qualitatively, the MLP-predicted effective conductivity of both validation rocks is nearly identical to the previously published solutions for simplified inclusion geometry and isotropic background conductivity (Figures~\ref{fig:validation-sphere}~and~\ref{fig:validation-ellipsoid}). This result confirms that the MLP approximations of the depolarization tensors, which are accurate to three significant digits on average, yield valid effective conductivity predictions. Quantitatively, the mean absolute percentage error between the predicted and actual effective conductivity is $0.08\pm 0.04$~\% for spherical inclusions and $0.04\pm 0.03$~\% for spheroidal inclusions. The error values and their uncertainties correspond to the mean and standard deviation across all frequencies, respectively.

\subsection{Anisotropic effective conductivity modeling}
This section demonstrates the use of the trained MLP in predicting the effective conductivity of rocks containing triaxial ellipsoidal inclusions in an anisotropic background. In these cases, analytical solutions allowing quantitative validation do not exist. Nevertheless, we demonstrate the consistency of the MLP predictions through examples of rock models that are conceptually simple to interpret. 

\subsubsection{Model A: Random triaxial ellipsoids with anisotropic background conductivity}

\begin{figure}[!htp]
\centering
\includegraphics[width=0.5\textwidth]{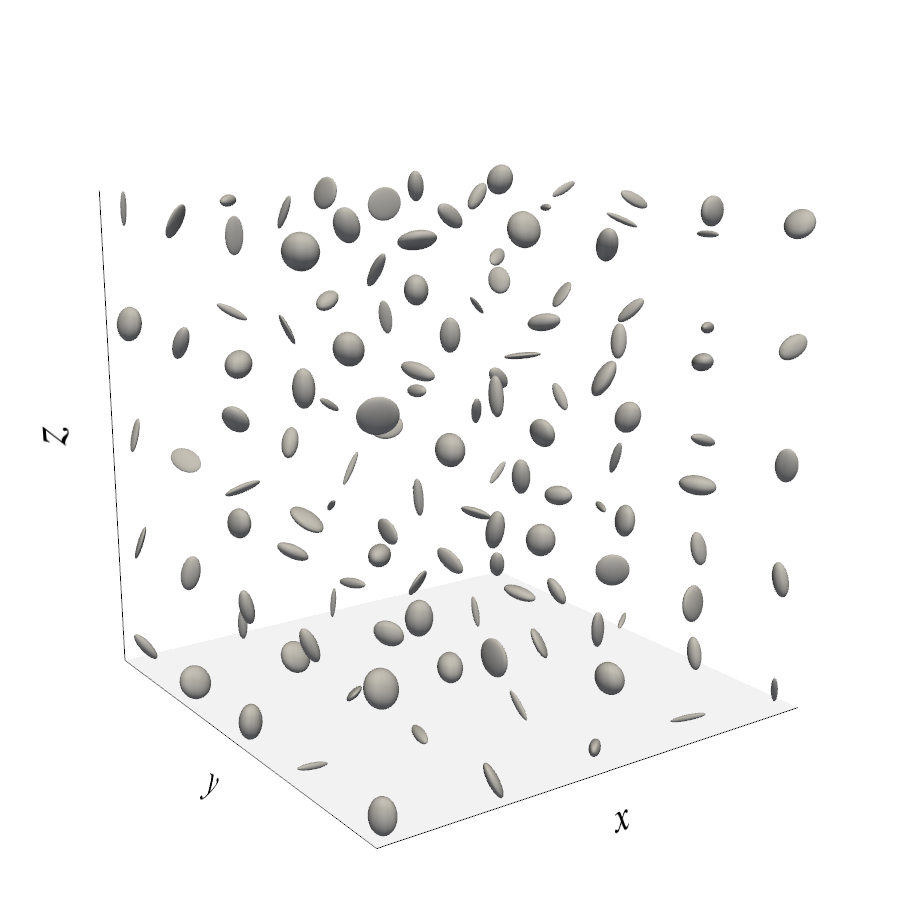}
\caption{Arbitrary unit volume of a rock containing randomly-oriented triaxial ellipsoidal inclusions in a background with anisotropic conductivity (Model A).}
\label{fig:ellipsoidsA}
\end{figure}

In this experiment, we generate a synthetic rock sample containing $10^6$ triaxial ellipsoids inclusions with random orientations. Figure~\ref{fig:ellipsoidsA} shows an arbitrary unit volume of rock model A for visualization purposes. 

\begin{figure}[!htp]
\centering
\includegraphics[width=0.5\textwidth]{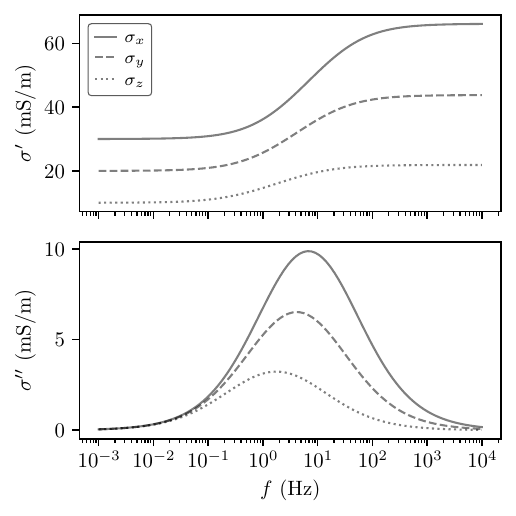}
\caption{MLP-predicted effective conductivity for a rock with random triaxial ellipsoidal inclusions in a background with anisotropic conductivity (Model A).}
\label{fig:model-A}
\end{figure}

For each inclusion in the rock sample of Figure~\ref{fig:ellipsoidsA}, the major axis ($a$) is set to 1 mm, and both semi-major and minor axes parameters (respectively $A$ and $B$) are uniformly distributed between 0.1 and 1. Furthermore, a random rotation matrix determines the orientation of each ellipsoid's major axis. The total volumetric fraction occupied by the inclusions in the medium is 20~\%, the intrinsic conductivity of the inclusions is $\sigma_l=10^4$~S/m, the $\varrho_l$ parameter is 0.8, and the empirical $\lambda_l$ parameter is 0.2~$\Omega$m$^2/$s$^{\varrho_l}$. With these properties, the medium should have isotropic effective conductivity due to the random orientations of the inclusions \citep{zhdanov_anisotropy_2008}. However, we use our formulation of the depolarization tensors to introduce an anisotropic background conductivity defined by $\sigma_{\mathrm{b},x}=0.03$, $\sigma_{\mathrm{b},y}=0.02$ and $\sigma_{\mathrm{b},z}=0.01$ S/m. Figure~\ref{fig:model-A} shows the MLP-predicted effective conductivity of rock model A.

It is evident from Figure~\ref{fig:model-A} that rock model A, with random ellipsoidal inclusions and anisotropic background conductivity, exhibits anisotropy in both the real and imaginary parts of its effective conductivity. As expected, the real conductivity in the low-frequency limit is consistent with that of the background conductivity. The maximum imaginary conductivities in the three directions also have contrasting intensities. Additionally, it is clear that as the directional background conductivity decreases, the critical frequency ($f_\mathrm{p}$) where the imaginary conductivity reaches its maximum shifts towards lower frequencies. When repeating the experiment multiple times, varying the inputs $C$ and $D$ for the neural network to simulate different anisotropy ratios of background conductivity, we find that if $\sigma_l \gg \sigma_\mathrm{b}$, 
\begin{equation}
    f_{\mathrm{p}, \{x,y,z\}} \propto \sigma_{\mathrm{b}, \{x,y,z\}}^{1/\varrho_l}.
\end{equation}

\subsubsection{Model B: Triaxial ellipsoids with isotropic background conductivity}

\begin{figure}[!htp]
\centering
\includegraphics[width=0.5\textwidth]{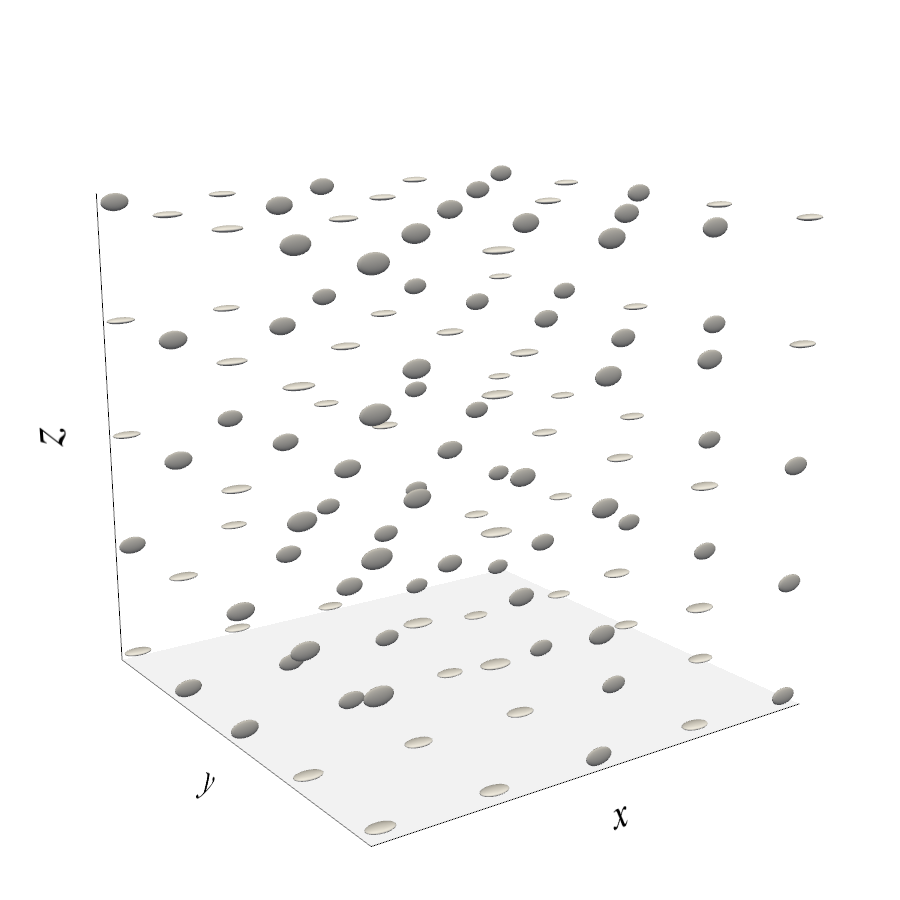}
\caption{Arbitrary unit volume of a rock with isotropic background conductivity containing two types of perpendicular triaxial ellipsoidal inclusions (Model B).}
\label{fig:ellipsoidsB}
\end{figure}

In this experiment, we generate a synthetic rock sample containing $10^6$ triaxial ellipsoidal inclusions with their major axis equal to 0.1~mm and aligned in the $x$ direction. The inclusions occupy a total volumetric fraction of 10~\% in an isotropic background medium which has a conductivity of $10^{-3}$~S/m. The inclusions are equally split into two groups. The first half have parameters $A=0.1$ and $B=0.6$, and the second half $A=0.6$ and $B=0.1$, meaning that the ellipsoids of both groups are flattened in perpendicular directions. Figure~\ref{fig:ellipsoidsB} shows an arbitrary unit volume of rock model B.  

\begin{figure}[!htp]
\centering
\includegraphics[width=0.5\textwidth]{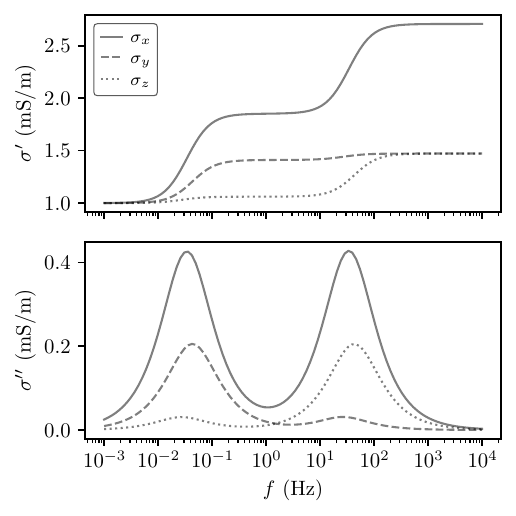}
\caption{MLP-predicted effective complex conductivity of a rock with isotropic background containing two types of perpendicular triaxial ellipsoids (Model B).}
\label{fig:model-B}
\end{figure}

The two inclusion groups share the following properties: $\sigma_l=5000$~S/m and $\varrho_l=1.0$. However, they have contrasting $\lambda_l$ parameters so that their critical frequencies do not overlap. The first inclusion group has $\lambda_l=10$~$\Omega$m$^2/$s$^{\varrho_l}$, and the second $\lambda_l=0.01$~$\Omega$m$^2/$s$^{\varrho_l}$. Figure~\ref{fig:model-B} shows the effective complex conductivity of rock model B in the $x$, $y$, and $z$ directions.

The effective conductivity of rock model B is, when measured in the three directions, vastly different (Figure~\ref{fig:model-B}). In the $x$ direction, there are two equal-intensity imaginary conductivity peaks at 0.03~Hz and 30~Hz. The $y$ direction shows one clear polarization peak from the second group of inclusions, with the first group's impact barely noticeable. Conversely, in the $z$ direction, the pattern is reversed. This illustrates the challenge in interpreting anisotropic media's effective conductivity based on rock properties and emphasizes the need for anisotropic IP models.

\subsection{Neural network analysis}
This section leverages our efficient neural network approach to investigate the GEMTIP model's interpretation, sensitivity, and error as a function of rock anisotropy.

\subsubsection{Anisotropic chargeability and inclusion fraction}
\begin{figure}[!htp]
\centering
\subfloat[]{%
  \includegraphics[clip,width=0.5\columnwidth]{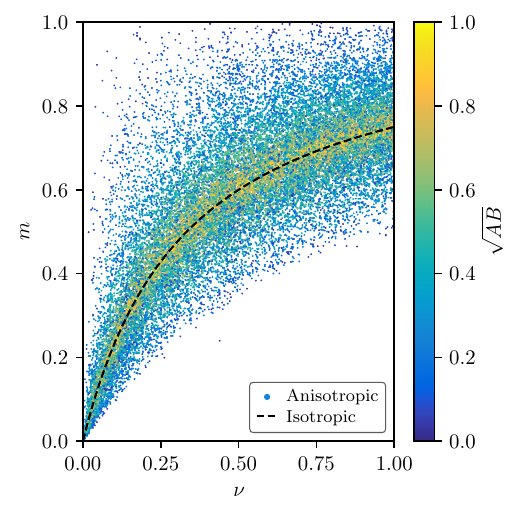}%
}
\subfloat[]{%
  \includegraphics[clip,width=0.5\columnwidth]{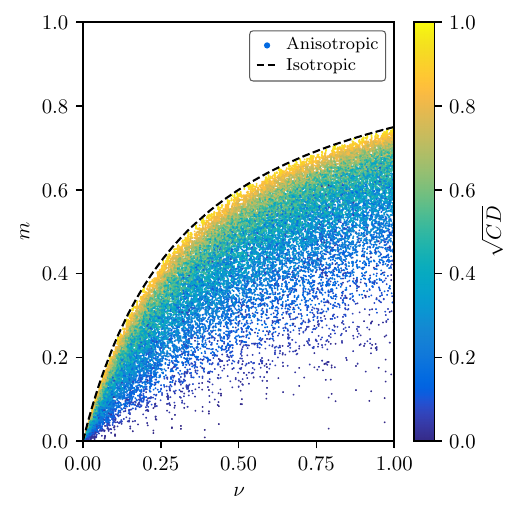}%
}
\caption{MLP-predicted chargeability ($m$) in the $x$, $y$ and $z$ directions as a function of the total volumetric fraction ($\nu$) of randomly-oriented polarizable inclusions for $10^4$ rocks. The dashed line is the analytical solution for isotropic media. (a) The ellipsoidal inclusion shape anisotropy parameters ($A$, $B$) vary. (b) The background conductivity anisotropy parameters ($C$, $D$) vary.}
\label{fig:shape-anisotropy-analysis}
\end{figure}

The direct relationship between chargeability and the volumetric fraction of polarizable inclusions is well-known. We analyze the impact of anisotropy on this relationship by simulating the effective conductivity of $10^4$ rocks containing ellipsoidal inclusions. We set $a_l=1$~mm, $\sigma_l=10^4$~S/m, $\varrho_l=1.0$, $\lambda_l=0.1$~$\Omega$m$^2/$s$^{\varrho_l}$, $\sigma_b=0.1$~S/m, and each rock's total volumetric fraction of inclusion $\nu$ is uniformly distributed between zero and one. Figure~\ref{fig:shape-anisotropy-analysis} shows the chargeability in the $x$, $y$ and $z$ directions as a function of $\nu$. In Figure~\ref{fig:shape-anisotropy-analysis}a, the background conductivity is isotropic and the semi-axes anisotropy parameters $A$ and $B$ vary between zero (indicative of thin sheets or rods) and one (spheres). In Figure~\ref{fig:shape-anisotropy-analysis}b, the inclusions are spherical and the conductivity anisotropy parameters $C$ and $D$ vary between zero and one.

As evidenced in Figure~\ref{fig:shape-anisotropy-analysis}, higher volumetric fractions of anisotropic inclusions are associated with higher chargeability values. The neural network accurately predicts the expected analytical relationship between chargeability and volumetric fraction when the inclusions are spherical (i.e., ${AB}\to 1$) and when the background conductivity is isotropic (i.e., ${CD}\to 1$). However, as the inclusion anisotropy increases (i.e., ${AB}\to 0$), the relationship between chargeability and volumetric content becomes diffuse (Figure~\ref{fig:shape-anisotropy-analysis}a). In other words, when the electric field is parallel to the smaller axes of the ellipsoids (see Figure~\ref{fig:model-B}), attenuation of the IP effect occurs in that direction and the volumetric content determination from chargeability is uncertain. In Figure~\ref{fig:shape-anisotropy-analysis}b,  increasing background conductivity anisotropy (i.e., ${CD}\to 0$) leads to underestimation of the chargeability values as a function of volumetric fraction.

\subsubsection{Anisotropic critical frequency and background conductivity}
\begin{figure}[!htp]
\centering
\subfloat[]{%
  \includegraphics[clip,width=0.5\columnwidth]{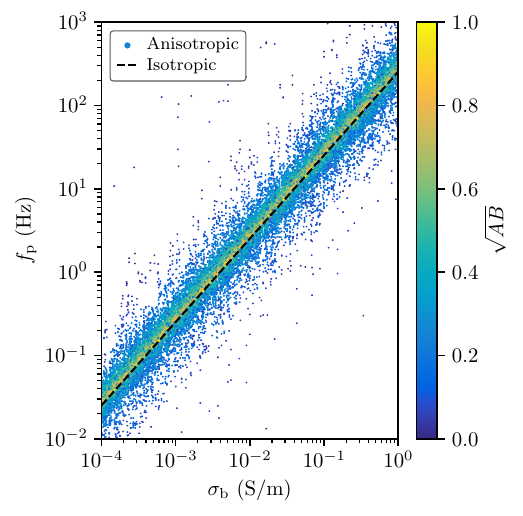}%
}
\subfloat[]{%
  \includegraphics[clip,width=0.5\columnwidth]{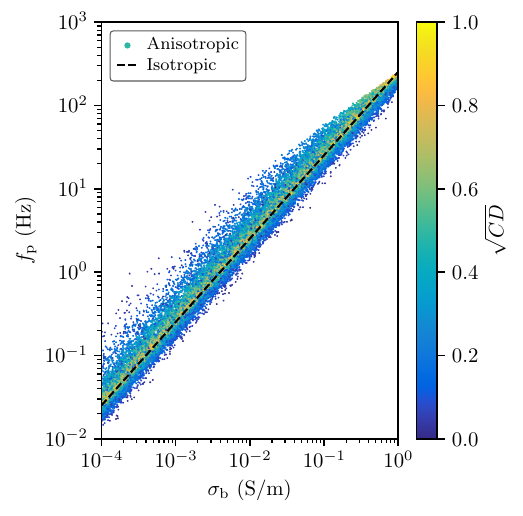}%
}
\caption{MLP-predicted critical polarization frequency ($f_\mathrm{p}$) in the $x$, $y$ and $z$ directions as a function of background conductivity ($\sigma_\mathrm{b}$) in those respective directions for $10^4$ rocks. The dashed line is the analytical solution for isotropic media. (a) The ellipsoidal inclusion shape anisotropy parameters ($A$, $B$) vary. (b) The background conductivity anisotropy parameters ($C$, $D$) vary.}
\label{fig:sigma-anisotropy-analysis}
\end{figure}

This section analyzes the relationship between the critical polarization frequency and the background conductivity of rocks. For this experiment, we set $a_l=1$~mm, $\nu_l=0.2$, $\sigma_l=10^4$~S/m, $\varrho_l=1.0$, $\lambda_l=1.0$~$\Omega$m$^2/$s$^{\varrho_l}$, and $\sigma_\mathrm{b,x}$ is log-uniformly distributed between $10^{-4}$ and 1~S/m. Figure~\ref{fig:sigma-anisotropy-analysis} shows $f_\mathrm{p}$ in the $x$, $y$ and $z$ directions as a function of $\sigma_\mathrm{b}$ for $10^4$ simulated rocks. In Figure~\ref{fig:sigma-anisotropy-analysis}a, the background conductivity is isotropic and the semi-axes anisotropy parameters $A$ and $B$ vary between zero (thin sheets or rods) and one (spheres). In Figure~\ref{fig:sigma-anisotropy-analysis}b, the inclusions are spherical and the conductivity anisotropy parameters $C$ and $D$ vary between zero and one.

As evidenced in Figure~\ref{fig:sigma-anisotropy-analysis}, increasing conductivity of the rock background leads to increasing critical polarization frequency. In particular, the neural network accurately predicts the expected analytical relationship between critical frequency and background conductivity when the background medium is isotropic (i.e., ${CD}\to 1$) and when the inclusions are spherical (i.e., ${AB}\to 1$). As the inclusion anisotropy increases (i.e., ${AB}\to 0$), the deviation from the expected relationship reaches up to two orders of magnitude (Figure~\ref{fig:sigma-anisotropy-analysis}a). Similarly, Figure~\ref{fig:sigma-anisotropy-analysis}b shows that the deviation from the expected relationship reaches up to one order of magnitude as the background conductivity anisotropy increases (i.e., ${CD}\to 0$). 

\subsubsection{Approximation error as a function of anisotropy}
\begin{figure}[!htp]
\centering
\includegraphics[width=0.5\textwidth]{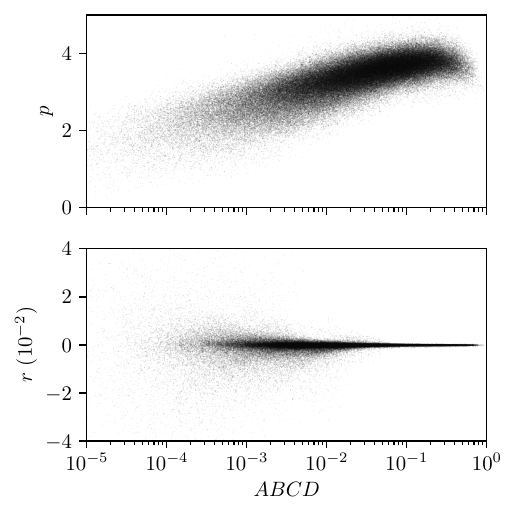}
\caption{Scatter plots of the $p$ and $r$ evaluation metrics against anisotropy parameters product $ABCD$. MLP accuracy decreases with increasing rock anisotropy.}
\label{fig:ABCD-error}
\end{figure}

The approximation errors of the depolarization tensor elements depend on the inclusion shape and host rock conductivity anisotropy, as demonstrated by the $p$ and $r$ metrics in relation to the product of anisotropy parameters $ABCD$ in Figure~\ref{fig:ABCD-error}.

Figure~\ref{fig:ABCD-error} shows that the MLP accurately predicts depolarization tensor elements with up to four significant digits when the $ABCD$ product of the inclusions exceeds 0.1. For $ABCD$ products near zero, the MLP's precision drops to an average of two significant digits. The impact of anisotropy on approximation error is also evident in the $r$ metric, where $ABCD$ products below 0.1 show high prediction spread, but products above 0.1 have $r$ values close to zero.

\subsubsection{Relative sensitivity indices}

Following the method of \cite{berube_bayesian_2023}, we evaluate the neural network's sensitivity by averaging its Jacobian matrix across $\mathcal{D}_\mathrm{t}$ and $\mathcal{D}_\mathrm{v}$. The relative sensitivity indices of the depolarization tensor elements to input parameters $A$, $B$, $C$, and $D$ are in Table~\ref{tab:sensitivity-indices}, where each row is normalized to sum to 100~\%.

\begin{table}[!ht]
\caption{Relative sensitivity indices (in \%) of the output depolarization tensor elements with respect to input anisotropy parameters $A$, $B$, $C$ and $D$.}
\centering
\begin{tabular}{lcccc}
\toprule
{} & {$A$} & {$B$} & {$C$} & {$D$} \\
\midrule
$\Gamma_{x}$ & $22.47 \pm 0.03$ & $22.54 \pm 0.03$ & $27.48 \pm 0.04$ & $27.51 \pm 0.04$ \\
$\Gamma_{y}$ & $27.66 \pm 0.03$ & $21.52 \pm 0.03$ & $21.19 \pm 0.03$ & $29.63 \pm 0.04$ \\
$\Gamma_{z}$ & $21.45 \pm 0.03$ & $27.76 \pm 0.03$ & $29.61 \pm 0.04$ & $21.19 \pm 0.03$ \\
$\Lambda_{x}$ & $31.37 \pm 0.04$ & $31.45 \pm 0.04$ & $18.60 \pm 0.03$ & $18.58 \pm 0.03$ \\
$\Lambda_{y}$ & $32.54 \pm 0.03$ & $32.86 \pm 0.04$ & $13.26 \pm 0.02$ & $21.34 \pm 0.03$ \\
$\Lambda_{z}$ & $32.78 \pm 0.04$ & $32.59 \pm 0.03$ & $21.36 \pm 0.03$ & $13.27 \pm 0.02$ \\
Average & $28.04 \pm 0.03$ & $28.12 \pm 0.03$ & $21.92 \pm 0.03$ & $21.92 \pm 0.03$ \\
\bottomrule
\end{tabular}
\label{tab:sensitivity-indices}
\end{table}

Table~\ref{tab:sensitivity-indices} shows that volume depolarization tensor elements have slightly higher sensitivity to background conductivity anisotropy than to ellipsoidal inclusions' shape anisotropy. However, for surface depolarization tensor elements, about one-third of the sensitivity comes from inclusion anisotropy, indicating a reversal in relative importance. Upon averaging the sensitivity of all depolarization tensor elements to input parameters, inclusion shape anisotropy emerges as more influential than background conductivity anisotropy in determining depolarization tensors. 


\subsubsection{Computation times}
\begin{figure}[!htp]
\centering
\includegraphics[width=0.5\textwidth]{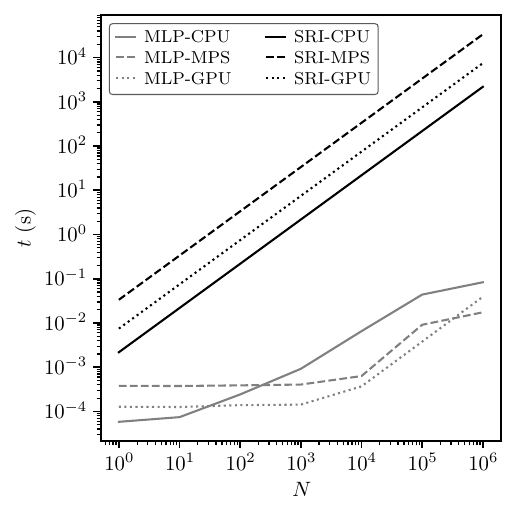}
\caption{Time ($t$) to solve the six depolarization tensor integrals for a rock containing $N$ polarizable inclusions with the MLP and SRI methods on the M1 Pro CPU, on the prior with MPS acceleration, and on the RTX~3060 GPU.}
\label{fig:timings}
\end{figure}

Computation time refers to the time required to evaluate all elements of the depolarization tensors. For the SRI method, this entails solving six integrals with $n$ evaluations per integrand. In contrast, the trained MLP predicts all six elements in a single operation through a forward pass, with the added benefit of vectorization enabling simultaneous integral evaluations for multiple inclusions. To compare MLP and SRI fairly, we chose $n=10^4$, ensuring a similar precision of four significant digits on the volume depolarization tensor elements for both methods (see Table~\ref{tab:comp-meth-niter}). We achieve consistent timings by performing each calculation 100 times and averaging the results. The computation times on the M1 Pro CPU, the prior with MPS acceleration, and the RTX~3060 GPU for up to $10^6$ inclusions are shown in Figure~\ref{fig:timings}.

Figure~\ref{fig:timings} gives a convincing argument regarding the numerical efficiency of the proposed MLP approach. Thanks to its inherent parallelization capabilities, the MLP predicts all six elements of the depolarization tensors for $10^6$ unique inclusions in under 0.1 s, whereas the computation times for the same task using SRI range from 1000~s to 30~000~s depending on the use of CPU, MPS or GPU.

\section{Conclusions}
This research streamlines and expedites IP modeling for anisotropic rocks. We achieve this goal in two parts: (1) by extending the GEMTIP model to incorporate triaxial ellipsoidal inclusions in anisotropic host rock, a notable departure from previous solutions limited to isotropic background conductivity and spheroidal grain shapes, and (2) by using a MLP neural network to approximate the depolarization tensors, offering a more efficient alternative to numerical integration methods. It may be possible to approximate the integrals using other methods, such as interpolation or open form solutions. However, it is inefficient to interpolate the integrals on general splines due to their dimensionality, and reformulating the problem as functions of elliptical integrals dependent on six anisotropy variables is a challenging endeavor. 

The quality and quantity of training data are factors that limit the accuracy of the MLP. Nevertheless, its ability to predict depolarization tensors with consistent precision and speed allows us to conduct new experiments on the anisotropy of IP effects. The relationships between critical frequency and background conductivity, as well as between chargeability and the volumetric fraction of polarizable inclusions, still hold for weakly anisotropic rocks. However, there are significant deviations from the expected relationships when interpreting IP data characteristic of highly anisotropic rocks. We also observe slight approximation errors in highly anisotropic rocks, which may indicate the need for further refinement of the MLP. In such cases, the depolarization tensor integrands are unstable, and the evaluation of their integrals could benefit from adaptive sampling techniques. Last, the relative sensitivity of depolarization tensors to inclusion and background anisotropy, at 56~\% and 44~\%, respectively, highlight the balanced influence of both factors on IP signals. 

Training the MLP is time-consuming but it only needs to be realized once. After training, the network performs IP simulations up to 100,000 faster than numerical integration without significant loss in accuracy. A notable use case of the GEMTIP model is predicting the effective conductivity of complex geological models parametrized by X-ray computer tomography, petrographic image analyses, or mineralogical statistics. Such simulations typically have prohibitive computing times for anisotropic rocks, but the MLP performs them in seconds on laptop computers. As a practical contribution, we release a pre-trained implementation of the MLP as an open-source Python package. This contribution opens new avenues for efficient rock-scale IP modeling, directly impacting data interpretation methods that aim to consider realistic and anisotropic geological scenarios. 

\begin{acknowledgments}
  C. L. Bérubé acknowledges funding from the FRQNT Research Support for New Academics under project titled \emph{Petrophysical modelling of the induced polarization effect with machine learning} (Grant No. 326054). J.-L. Gagnon is supported by a NSERC Undergraduate Student Research Award. Special thanks are extended to Dr. Frédérique Baron for advice on data visualization and insightful manuscript revisions. 
  
\end{acknowledgments}

\bibliographystyle{seg}  
\bibliography{references}

\clearpage
\append{Anisotropic depolarization tensors}
\label{sect:supportTens}

Here we provide additional details on the derivation of the depolarization tensor integrals in the case of diagonal tensor conductivity and triaxial ellipsoidal inclusions.

 \subsection{Chain derivation of the general Green function}
Using the chain derivation rule, the gradient of the general Green's function reads
\begin{equation}
 \nabla_{x,y,z} G (X,Y,Z) = \derip{G_{x,y,z}}{X}\derip{X}{x}\,\mathbf{\hat{x}} + \derip{G_{x,y,z}}{Y}\derip{Y}{y}\,\mathbf{\hat{y}} + \derip{G_{x,y,z}}{Z}\derip{Z}{z}\,\mathbf{\hat{z}}.
\end{equation}
The usual gradient, but with respect to $X,Y,Z$ rather than $x, y, z$ is obtained by introducing a transformation $\mathbf{T}$, such that
\begin{equation}
   \nabla_{x,y,z} G = \overds{T}\nabla_{X,Y,Z} G,
\end{equation}
where 
\begin{equation}
 \overds{T} = \begin{pmatrix}
\derip{X}{x} & 0 & 0 \\
0 & \derip{Y}{y} & 0\\
0 & 0 & \derip{Z}{z}
\end{pmatrix} = \begin{pmatrix}
   \sigma_{\mathrm{b},x} & 0 & 0 \\
   0 & \sigma_{\mathrm{b},y} & 0\\
   0 & 0 & \sigma_{\mathrm{b},z}
    \end{pmatrix}^{-1/2}.
\end{equation}

 \subsection{Volume depolarization tensor}
 The norm of the vectors $\mathbf{r}'$, $\mathbf{R}'$ and $\mathbf{n}'$ are 
\begin{align}
       |\mathbf{r'}| &= \sqrt{a^{2}\sin^2(\theta)\cos^2(\phi) + b^{2}\sin^2(\theta)\sin^2(\phi) + c^{2}\cos^2(\theta)}, \\
       |\mathbf{R'}| &= \sqrt{a'^{2}\sin^2(\theta)\cos^2(\phi) + b'^{2}\sin^2(\theta)\sin^2(\phi) + c'^{2}\cos^2(\theta)}, \\
  |\mathbf{n'}| &= \sqrt{a^{-2}\sin^2(\theta)\cos^2(\phi) + b^{-2}\sin^2(\theta)\sin^2(\phi) + c^{-2}\cos^2(\theta)}.
\end{align}
Consequently, the product $\mathbf{n'}\mathbf{R'}$ in the volume depolarization tensor expression is 
\begin{equation}
   \mathbf{n'}\mathbf{R'}= \begin{pmatrix}
   \frac{1}{\sqrt{\sigma_{\mathrm{b},x}}}\sin^2(\theta)\cos^2(\phi) & \frac{a'}{b}\sin^2(\theta)\cos(\phi)\sin(\phi) & \frac{a'}{c}\sin(\theta)\cos(\phi)\cos(\theta) \\
   \frac{b'}{a}\sin^2(\theta)\cos(\phi)\sin(\phi) & \frac{1}{\sqrt{\sigma_{\mathrm{b},y}}}\sin^2(\theta)\sin^2(\phi) & \frac{b'}{c}\sin(\theta)\sin(\phi)\cos(\theta)\\
    \frac{c'}{a}\sin(\theta)\cos(\phi)\cos(\theta)& \frac{c'}{b}\sin(\theta)\sin(\phi)\cos(\theta) & \frac{1}{\sqrt{\sigma_{\mathrm{b},z}}}\cos^2(\theta)
\end{pmatrix}.
\end{equation}
Using Equation~\ref{eq:partInt}, the volume depolarization tensor of inclusion $l$ thus reads
\begin{equation}
    \overds{\Gamma}_{l} = -\overds{T}\int_{0}^{2\pi}\int_{0}^{\pi}\frac{\mathrm{d}S}{4\pi \sigma_{s}|\mathbf{R}'|^3|\mathbf{n}'|}\mathbf{n'}\mathbf{R'}.
\end{equation}

The next step is developing the surface element $\mathrm{d}S$ from the vector product of $\mathbf{r'}_{\theta}$ and $\mathbf{r'}_{\phi}$. It is important to note that even under the variable change $\mathbf{R'}$, the integral is not calculated over the modified ellipsoid, but on the normal ellipsoid, thus the surface element is a function of $a,b,c$ and reads
\begin{align}
\begin{split}
    \mathrm{d}S &= \frac{\sqrt{a^2b^2\sin^2(2\theta) + 4a^2c^2\sin^2(\phi)\sin^4(\theta) + 4b^2c^2\sin^4(\theta)\cos^2(\phi)}}{2}\,\mathrm{d}\theta\mathrm{d}\phi \\
    &= |\sin{\theta}| \sqrt{c^{2}b^2\sin^2(\theta)\cos^2(\phi) + a^{2}c^{2}\sin^2(\theta)\sin^2(\phi) + a^{2}b^{2}\cos^2(\theta)}\,\mathrm{d}\theta \mathrm{d}\phi \\
    &= abc |\sin{\theta}| |\mathbf{n}'| \,\mathrm{d}\theta \mathrm{d}\phi.
\end{split}
\end{align}
Finally, by including $\mathrm{d}S$ in the volume depolarization tensor equation, we get
\begin{equation}
    \overds{\Gamma}_{l} = -\frac{abc}{4\pi\sigma_{s}}\overds{T}\int_{0}^{2\pi}\int_{0}^{\pi}\frac{\mathrm{d}\theta \mathrm{d}\phi\, |\sin{\theta}|}{|\mathbf{R}'|^3}\mathbf{n'}\mathbf{R'},
\label{eq:TensDipEll3App}
\end{equation}
for which, after \cite{milton_2002_theory}, the solution involves elliptic functions of the first kind ($F$) and of the second kind ($E$) with 
\begin{align}
\begin{split}
    \Gamma_{x} &= \frac{abc(F(k',\varphi')-E(k',\varphi'))}{{\sigma}_{s}\sigma_{\mathrm{b},x}(a'^2-b'^2)\sqrt{a'^2-c'^2}},\\
    \Gamma_{y} &= \frac{abc}{{\sigma}_{s}\sigma_{\mathrm{b},y}}\left(\frac{(F(k',\varphi')-E(k',\varphi'))}{(a'^2-b'^2)\sqrt{a'^2-c'^2}} + \frac{E(k',\varphi')}{(b'^2-c'^2)\sqrt{a'^2-c'^2}}-\frac{c'}{a'b'(b'^2-c'^2)}\right),\\
    \Gamma_{z} &= \frac{abc}{\sigma_{s}\sigma_{\mathrm{b},x}}\left(-\frac{E(k',\varphi')}{(b'^2-c'^2)\sqrt{a'^2-c'^2}}+\frac{b'}{a'c'(b'^2-c'^2)}\right),
    \label{eq:depFactEllApp}
\end{split}
\end{align}
provided that $a' > b' > c'$.
The parameters of the elliptic functions are the phase $\varphi' = \sqrt{\arcsin\left({1-\frac{c'^2}{a'^2}}\right)}$ and the modulus $k' = \sqrt{\frac{a'^2-b'^2}{a'^2-c'^2}}$.

We verify the consistency of the anisotropic volume depolarization tensor formulation by simplifying it to the isotropic case. Assuming isotropic conductivity, i.e., $\sigma_{\mathrm{b},x},\sigma_{\mathrm{b},y},\sigma_{\mathrm{b},z}=\sigma_\mathrm{b}$, the expressions of $\mathrm{d}S$, $|\mathbf{R}'|$, $|\mathbf{n}'|'$ and $\sigma_{s}$ simplify and we get
\begin{align}
\begin{split}
    \overds{\Gamma}_{l} &= -\frac{abc}{4\pi\sigma_\mathrm{b}^{\sfrac{3}{2}}}\int_{0}^{2\pi}\int_{0}^{\pi}\frac{\mathrm{d}\theta \mathrm{d}\phi\,|\sin{(\theta})|}{|\mathbf{R}'|^3}\overds{T}\left(\mathbf{n'}\mathbf{R'}\right) \\
    &= -\frac{abc\cdot\sigma^{\sfrac{3}{2}}_\mathrm{b}}{4\pi\sigma_\mathrm{b}^{\sfrac{3}{2}} \sigma_\mathrm{b}^{\sfrac{1}{2}}}\int_{0}^{2\pi}\int_{0}^{\pi}\frac{\mathrm{d}\theta \mathrm{d}\phi|\sin{(\theta})|}{|\mathbf{r}'|^3}\mathbf{n'}\mathbf{R'}\\
    &= -\frac{abc}{4\pi\sigma_\mathrm{b}^{\sfrac{1}{2}}\sigma_\mathrm{b}^{\sfrac{1}{2}}}\int_{0}^{2\pi}\int_{0}^{\pi}\frac{\mathrm{d}\theta \mathrm{d}\phi|\sin{(\theta})|}{|\mathbf{r}'|^3}\mathbf{n'}\mathbf{r'}\\
    &= -\frac{abc}{4\pi\sigma_\mathrm{b}}\int_{0}^{2\pi}\int_{0}^{\pi}\frac{\mathrm{d}\theta \mathrm{d}\phi\sin{(\theta})}{|\mathbf{r}'|^3}\mathbf{n'}\mathbf{r'},
\end{split}
\end{align}
where $\mathbf{n'}\mathbf{r'}$ is 
\begin{equation*}
   \mathbf{n'}\mathbf{r'} = \begin{pmatrix}
  \sin^2(\theta)\cos^2(\phi) & \frac{a}{b}\sin^2(\theta)\cos(\phi)\sin(\phi) & \frac{a}{c}\sin(\theta)\cos(\phi)\cos(\theta) \\
   \frac{b}{a}\sin^2(\theta)\cos(\phi)\sin(\phi) & \sin^2(\theta)\sin^2(\phi) & \frac{b}{c}\sin(\theta)\sin(\phi)\cos(\theta)\\
    \frac{c}{a}\sin(\theta)\cos(\phi)\cos(\theta)& \frac{c}{b}\sin(\theta)\sin(\phi)\cos(\theta) & \cos^2(\theta)
\end{pmatrix},
\end{equation*}
and $\overds{\Gamma}_{l}$ is equivalent to the solution of \cite{zhdanov_generalized_2008} for spherical inclusions.

\subsection{Surface depolarization tensor}
We start with
\begin{align}
\begin{split}
    \nabla'\left(\frac{\mathbf{R'}}{|\mathbf{R'}|^{3}}\right) &= \mathbf{R'} \nabla' |\mathbf{R}'|^{-3} + |\mathbf{R}'|^{-3}\nabla'\mathbf{R'} \\
    &= \left(-3 \mathbf{R'} \frac{\mathbf{R'}}{|\mathbf{R}'|^{5}} + \frac{\unitary{x}\unitary{x} + \unitary{y}\unitary{y} + \unitary{z}\unitary{z}}{|\mathbf{R}'|^{3}}\right)\overds{T}\\
    &= \left(\frac{-3}{|\mathbf{R}'|^{5}} (\mathbf{R'}\mathbf{R'}) + \frac{\overds{I}}{|\mathbf{R}'|^{3}}\right)\overds{T}.
    \label{eq:GreenDyad}
\end{split}
\end{align}
Using Equation~\ref{eq:gradSurf}, the tensor products in the surface depolarization tensor are
\begin{align}
        \mathbf{n'}\mathbf{n'} &= \begin{pmatrix}
    \frac{1}{a^2}\sin^2(\theta)\cos^2(\phi) & \frac{1}{ab}\sin^2(\theta)\cos(\phi)\sin(\phi) & \frac{1}{ac}\sin(\theta)\cos(\phi)\cos(\theta) \\
   \frac{1}{ab}\sin^2(\theta)\cos(\phi)\sin(\phi) &  \frac{1}{b^2}\sin^2(\theta)\sin^2(\phi) & \frac{1}{bc}\sin(\theta)\sin(\phi)\cos(\theta)\\
    \frac{1}{ac}\sin(\theta)\cos(\phi)\cos(\theta)& \frac{1}{bc}\sin(\theta)\sin(\phi)\cos(\theta) &  \frac{1}{c^2}\cos^2(\theta)
    \end{pmatrix},
\end{align}
and
\begin{align}
   \mathbf{R'}\mathbf{R'} &= \begin{pmatrix}
    a'^2\sin^2(\theta)\cos^2(\phi) & a'b'\sin^2(\theta)\cos(\phi)\sin(\phi) & a'c'\sin(\theta)\cos(\phi)\cos(\theta) \\
   a'b'\sin^2(\theta)\cos(\phi)\sin(\phi) &  b'^2\sin^2(\theta)\sin^2(\phi) & b'c'\sin(\theta)\sin(\phi)\cos(\theta)\\
    a'c'\sin(\theta)\cos(\phi)\cos(\theta)& b'c'\sin(\theta)\sin(\phi)\cos(\theta) &  c'^2\cos^2(\theta)
    \end{pmatrix}.
\end{align}
We can then write the full expression for the surface depolarization tensor as
\begin{align}
\begin{split}
    \overds{\Lambda}_{l} &= -\int_{0}^{2\pi}\int_{0}^{\pi}\frac{\mathrm{d}S}{4\pi \sigma_{s}|\mathbf{R}'|^5|\mathbf{n}'|^2}\underbrace{\left(-3(\mathbf{R'}\mathbf{R'}) + |\mathbf{R}'|^{2}\overds{I}\right)\mathbf{n'}\mathbf{n'}\left(\overds{T}\right)^{2}}_{\overds{Q'}},\\
    &= -\frac{abc}{4\pi\sigma_{s}}\int_{0}^{2\pi}\int_{0}^{\pi}\frac{\mathrm{d}\theta \mathrm{d}\phi\, \sin{\theta}}{|\mathbf{R}'|^5|\mathbf{n}'|}\overds{Q'},
\label{eq:TensDipEll2App}
\end{split}
\end{align}
where the diagonal elements of $\overds{Q'}$ are 
\begin{align}
\begin{split}
    Q_{x} &= \sin^2{(\theta)}\cos^2{(\phi)}\left(-3\eta'_x + \frac{|\mathbf{R}'|^{2}}{a^{2}} \right)\frac{1}{\sigma_{\mathrm{b},x}},\\
    Q_{y} &= \sin^2{(\theta)}\sin^2{(\phi)}\left(-3\eta'_y + \frac{|\mathbf{R}'|^{2}}{b^{2}} \right)\frac{1}{\sigma_{\mathrm{b},y}},\\
    Q_{z} &= \cos^2{(\theta)}\left(-3\eta'_z + \frac{|\mathbf{R}'|^{2}}{c^{2}} \right)\frac{1}{\sigma_{\mathrm{b},z}} \label{eq:Qfactor},
\end{split}
\end{align}
and where
\begin{align}
\begin{split}
    \eta'_\iota &= (\sigma_{\mathrm{b},x}\sigma_{\mathrm{b},\iota})^{-1/2}\sin^2(\theta)\cos^2(\phi) \\
    &\qquad + (\sigma_{\mathrm{b},y}\sigma_{\mathrm{b},\iota})^{-1/2}\sin^2(\theta)\sin^2(\phi) \\
    &\qquad + (\sigma_{\mathrm{b},z}\sigma_{\mathrm{b},\iota})^{-1/2}\cos^2(\theta),
\end{split}
\end{align}
with indices $\iota=\{x,y,z\}$.

We verify the consistency of the surface depolarization tensor by considering isotropic conductivity, i.e., $\sigma_{\mathrm{b},x},\sigma_{\mathrm{b},y},\sigma_{\mathrm{b},z}=\sigma_\mathrm{b}$, thus simplifying its expression to 
\begin{align}
\begin{split}
    \overds{\Lambda}_{l} &= -\frac{abc}{4\pi\sigma_\mathrm{b}^{\sfrac{5}{2}}}\int_{0}^{2\pi}\int_{0}^{\pi}\frac{\mathrm{d}\theta \mathrm{d}\phi\, \sin{\theta}}{|\mathbf{R}'|^5|\mathbf{n}'|}\overds{Q}'\\
    &= -\frac{abc\sigma_\mathrm{b}^{\sfrac{5}{2}}}{4\pi\sigma_\mathrm{b}^{\sfrac{5}{2}}}\int_{0}^{2\pi}\int_{0}^{\pi}\frac{\mathrm{d}\theta \mathrm{d}\phi\, \sin{\theta}}{|\mathbf{r}'|^5|\mathbf{n}'|}\overds{Q}'\\
    &= -\frac{abc}{4\pi\sigma_\mathrm{b}}\int_{0}^{2\pi}\int_{0}^{\pi}\frac{\mathrm{d}\theta \mathrm{d}\phi\, \sin{\theta}}{|\mathbf{r}'|^5|\mathbf{n}'|}\overds{Q},
\end{split}
\end{align}
where
\begin{equation}
    \overdv{Q} = \left(-3\mathbf{r'}\mathbf{r'} + |\mathbf{r}'|^{2}\overds{I}\right)\mathbf{n'}\mathbf{n'}.
\end{equation}
Finally, the diagonal elements of $\overdv{Q}$ are
\begin{align}
\begin{split}
    Q_{x} &= \sin^2{(\theta)}\cos^2{(\phi)}\left(-3+ \frac{|\mathbf{r}'|^{2}}{a^{2}} \right)\\
    Q_{y} &= \sin^2{(\theta)}\sin^2{(\phi)}\left(-3+ \frac{|\mathbf{r}'|^{2}}{b^{2}} \right)\\
    Q_{z} &= \cos^2{(\theta)}\left(-3+ \frac{|\mathbf{r}'|^{2}}{c^{2}} \right),
\label{eq:QfactorApp}
\end{split}
\end{align}
and $\overds{\Lambda}_{l}$ is equivalent to the solution of \cite{zhdanov_generalized_2008} for spherical inclusions.

\end{document}